\documentclass{aastex}
\usepackage{emulateapj5}

\submitted{Submitted 17 November 2003; Accepted 2 January 2004 -- To appear in the April, 2004 AJ}
\newcommand{\etal}{{\it et\thinspace al.}\ }
\newcommand{\kms}{km\thinspace s$^{-1}$}
\newcommand{\simlt}{\ {\raise-.5ex\hbox{$\buildrel<\over\sim$}}\ }

\lefthead{Gronwall \etal}
\righthead{KPNO International Spectroscopic Survey IV.}

\begin{document}

\title{The KPNO International Spectroscopic Survey.  \\ IV. H$\alpha$-selected Survey List 2.}

\author{Caryl Gronwall\altaffilmark{1,2}, John J. Salzer\altaffilmark{1}, Vicki L. Sarajedini\altaffilmark{3}, Anna Jangren, and Laura Chomiuk\altaffilmark{4} }
\affil{Astronomy Department, Wesleyan University, Middletown, CT 06459; 
slaz@astro.wesleyan.edu, anna@astro.wesleyan.edu}

\author{J. Ward Moody}
\affil{Department of Physics \& Astronomy, Brigham Young University, Provo, UT 84602; jmoody@astro.byu.edu}

\author{Lisa M. Frattare\altaffilmark{1}}
\affil{Space Telescope Science Institute, Baltimore, MD 21218; frattare@stsci.edu}

\author{Todd A. Boroson}
\affil{National Optical Astronomy Obs., P.O. Box 26732, Tucson, AZ 85726; tyb@noao.edu}

\altaffiltext{1}{Visiting Astronomer, Kitt Peak National Observatory, National Optical 
Astronomy Observatory, which is operated by the Association of Universities 
for Research in Astronomy, Inc. (AURA) under cooperative agreement with the 
National Science Foundation.} 
\altaffiltext{2}{present address: Department of  Astronomy \& Astrophysics, Pennsylvania State University, University Park, PA 16802; caryl@astro.psu.edu.}
\altaffiltext{3}{present address: Astronomy Department, University of Florida,
Gainesville, FL, 32611; vicki@astro.ufl.edu.}
\altaffiltext{4}{present address: UCO/Lick Observatory, UC Santa Cruz,
Santa Cruz, CA, 95064; laura@astro.ucsc.edu.}
%\clearpage

\begin{abstract}
The KPNO International Spectroscopic Survey (KISS) is an objective-prism
survey for extragalactic emission-line objects.  It combines many of the 
features of previous slitless spectroscopic surveys with the advantages of 
modern CCD detectors, and is the first purely digital objective-prism survey
for emission-line galaxies.  Here we present the second list of emission-line 
galaxy candidates selected from our red spectral data, which cover the wavelength
range 6400 to 7200 \AA.  In most cases, the detected emission line is 
H$\alpha$.  The current survey list covers a 1.6-degree-wide strip located 
at $\delta$(1950) = 43$\arcdeg$~30$\arcmin$ and spans the RA range 
11$^h$~55$^m$ to 16$^h$~15$^m$.  The survey strip runs through the center
of the Bo\"otes Void, and has enough depth to adequately sample the far side of
the void. An area of 65.8 deg$^2$ is covered.  A total of 1029 candidate emission-line 
objects have been selected for inclusion in the survey list (15.6 per deg$^2$).  We 
tabulate accurate coordinates and photometry for each source, as well as estimates 
of the redshift and emission-line flux and equivalent width based on measurements 
of the digital objective-prism spectra.  The properties of the KISS emission-line
galaxies are examined using the available observational data.  Although the
current survey covers only a modest fraction of the total volume of the Bo\"otes Void, 
we catalog at least twelve objects that appear to be located within the void.
Only one of these objects has been recognized previously as a void galaxy.
\end{abstract}

% The different journals have different requirements for keywords.  The
% keywords.apj file, found on aas.org in the pubs/aastex-misc directory, 
% contains a list of keywords used with the ApJ and Letters.  These are 
% usually assigned by the editor, but authors may include them in their 
% manuscripts if they wish. 


%************************************************************************

\section{Introduction}

Objective-prism surveys for extragalactic emission-line and/or UV-excess objects 
have been a fruitful and efficient method for cataloging galaxies with high levels
of activity (both star formation and active nuclei) for many years.  Most surveys
of this kind have utilized photographic plates  as their detectors (e.g., Markarian 
1967; Smith, Aguirre, \& Zemelman 1976; MacAlpine, Smith, \& Lewis 1977; Pesch \& 
Sanduleak 1983; Wasilewski 1983;  Markarian, Lipovetskii, \& Stepanian 1983; Zamorano 
\etal 1994; Popescu \etal 1996; Surace \& Comte 1998; Ugryumov \etal 1999).  
			
Despite their obvious advantages as detectors, CCDs were slow in replacing plates
on Schmidt telescopes due to their small areal coverage.  However, the availability of 
large-format CCDs in the early- and mid-1990s prompted many observatories to switch to 
the use of CCDs.  This change in turn prompted a group of astronomers interested in
extragalactic emission-line surveys to initiate a new CCD-based objective-prism survey.  
Called the KPNO International Spectroscopic Survey (KISS), it is the first fully digital 
objective-prism survey for emission-line galaxies (ELGs).
 
The goal of KISS  is to survey a substantial area of the sky (300--400 sq. deg.) to flux levels
substantially deeper than previous photographic surveys.  When completed, we expect to
have cataloged  roughly 5000 ELGs.  Because of the great depth of the survey, we do not
need to cover as much area as previous surveys to be able to generate large
samples of active galaxies for subsequent study.  For example, the Markarian
survey covered roughly 15,000 sq. deg. and catalogued 1500 objects.  In the first two red
KISS catalogs we have detected 2157 objects in only 128 sq. deg. Another important factor 
is that, due to the digital nature of the survey data, the completeness characteristics 
and flux limits are readily measurable {\it from the survey data themselves}. Hence KISS 
is uniquely suited for a broad range of studies that require statistically complete samples 
of galaxies.

This is the fourth paper in the KISS series.   The first presents a complete description of the
survey method, including a discussion of the survey data and its associated uncertainties  
(Salzer \etal 2000; hereafter Paper I).  The first survey list of H$\alpha$-selected ELGs,
informally referred to as the red survey, is given in Salzer \etal (2001; hereafter KR1), while 
the first list of [\ion{O}{3}]-selected galaxies (the blue survey) is found in Salzer \etal (2002; 
hereafter KB1).   The current paper follows a format similar to KR1; for the sake of brevity,
the reader is referred to Paper I and KR1 for many details.  The  observational data and
image processing are described in Section 2, while the new list of ELG candidates
is presented in Section 3.  The properties of the new list of H$\alpha$-selected ELGs are
described in Section 4, while our results are summarized in Section 5.

%************************************************************************

\section{Observations \& Reductions}

All survey data were acquired using the 0.61-meter Burrell Schmidt 
telescope\footnote{Observations made with the Burrell Schmidt telescope of 
the Warner and Swasey Observatory, Case Western Reserve University.}. 
The detector used for all data reported here was a 2048 $\times$ 4096 
pixel SiTe CCD.   This CCD is not identical to the one used for the previous
two survey lists, giving a different image scale and field-of-view.  The CCD has 
15-$\micron$ pixels, yielding an image 
scale of 1.43 arcsec/pixel at the Newtonian focus of the telescope. The overall 
field-of-view was 50 $\times$ 100 arcmin, and each image covered 1.37 square 
degrees.  The long dimension of the CCD was oriented north-south during our survey 
observations.  The red survey spectral data were obtained with a 4$\arcdeg$ 
prism, which provided a reciprocal dispersion of 17 \AA/pixel at H$\alpha$ with the
new CCD.   The spectral data were obtained through a special filter designed for
the survey, which covered the spectral range 6400 -- 7200 \AA\ (see Figure 1 of 
Paper I for the filter transmission curve).

The current survey was carried out in a strip at constant declination by observing a
series of contiguous fields offset from each other in right ascension.  The survey
is centered  at $\delta$(1950) = 43$\arcdeg$~30$\arcmin$, and each field covers
100 arcmin (1.6 degrees) in the declination direction.   The RA coverage of the survey
strip is from 11$^h$~55$^m$ to 16$^h$~15$^m$.  The location of the strip was chosen
to pass through the center of the Bo\"otes Void (Kirshner \etal 1981, 1983, 1987).  
We selected this area of the sky because of our interest in the spatial distribution 
of galaxies in and around the void.  Additionally, there have been a number of previous 
surveys that have looked at this region of the sky (Sanduleak \& Pesch 1982, 1987; 
Moody \etal 1987; Aldering 1990), giving us a comparison sample of previous 
objective-prism-selected and line-selected galaxies.

As with our previous survey strips, we obtained images of each survey field both
with and without the objective prism on the telescope.  The images taken without
the prism (referred to as direct images) were obtained through standard B and V
filters.  The direct images were photometrically calibrated, and provide accurate 
astrometry and photometry for all sources in the survey fields.  We used uniform 
exposure times for all survey fields: 4 $\times$ 720 s for the objective-prism (spectral)
data, and 2 $\times$ 300 s for V and 1 $\times$ 600 s in B for the direct images.
The telescope was dithered by a small amount ($\sim$10 arcsec) between exposures.

%\placetable{table:tab1}

Table \ref{table:tab1} lists the observing runs during which the current set of survey
fields were observed. The first column gives the UT dates of the run, while the second 
column indicates the number of nights on which observations were obtained.  At least 
some data were obtained on 26 of 27 scheduled nights (96\%).  The last two columns
indicate the number of direct and spectral images, respectively, obtained during each run. 
It was common practice to observe in both direct and spectral modes during parts of
each run, although it was not always the case that the direct and spectral images of a
given field were obtained during the same run.
 
All data reduction took place using the Image Reduction and Analysis Facility 
(IRAF\footnote{IRAF is distributed by the National Optical Astronomy Observatory, 
which is operated by the Association of Universities for Research in Astronomy, Inc., 
(AURA) under cooperative agreement with the National Science Foundation.}) software.  
A special package of IRAF-based routines that were written by members of the KISS team 
was used for most of the data analysis.
% This package is described in Herrero \etal (2003).
Full details of the observing procedures and data reduction methods are given in
Paper I and KR1.  

%************************************************************************

\section{List 2 of the KPNO International Spectroscopic Survey}

\subsection{Selection Criteria}

The selection of the second red (H$\alpha$) list of ELG candidates was carried
out in precisely the same fashion as the original red list (KR1).  Full details are
presented in Paper I and KR1.  To briefly summarize, we use our automated KISS software to 
evaluate the extracted objective-prism spectrum of each object located within a
survey field.  All objects with spectral features that rise more than five times the
local noise above the continuum level are flagged as potential ELGs.  This 5$\sigma$ 
threshold is the primary selection criterion of the survey, and was arrived at after
substantial testing during the early phases of the KISS project.  Following the
initial automated selection, all candidates are visually examined, and spurious
sources are removed from the sample.  Finally, the objective-prism images are
scanned visually for sources that might have been missed by the software.  These
tend to be objects where the emission line is redshifted to the red end of the
objective-prism spectrum, so that the software cannot detect continuum on both
sides of the line.  The combination of our automated selection process and our
careful visual checking helps to ensure a high degree of reliability that the KISS
ELG candidates are real, and that the sample is largely complete for all objects
with 5$\sigma$ emission lines.

As described in KR1, we also flag objects that have emission lines between
4$\sigma$ and 5$\sigma$ during our selection process.  These 4$\sigma$ detections
represent objects with somewhat weaker emission lines than the main KISS sample, but 
they are nonetheless valid ELG candidates and likely include a number of interesting
sources.  However, these objects do not constitute a statistically complete sample
in the same sense as the main ($>$ 5$\sigma$) list.  We report the 4--5$\sigma$
sources in a secondary list of ELG candidates (see Appendix), which should be
thought of as a supplement to the main KISS catalog.

\subsection{The Survey}

The list of ELG candidates selected in the second red survey is presented 
in Table \ref{table:tab2}.  Because the survey data includes both spectral 
images and photometrically-calibrated direct images, we are 
able to include a great deal of useful information about each source, such  
as accurate photometry and astrometry and estimates of the redshift, 
emission-line flux and equivalent width.  Only the first page of the
table is printed here; the complete table is available in the electronic
version of this paper.

%\placetable{table:tab2}

The contents of the survey table are as follows.  Column 1 gives a running 
number for each object in the survey with the designation KISSR $xxxx$, where 
KISSR stands for ``KISS red'' survey.  This is to distinguish it from the blue 
KISS survey (KB1).  The KR1 survey list included KISSR objects 1--1128, 
and here we present KISSR objects 1129--2157.  
Columns 2 and 3 give the object identification from the KISS database tables, 
where the first number indicates the survey field (F$xxxx$), and the second number 
is the identification number within the table for that galaxy. This identifier is 
necessary for locating the KISS ELGs within the survey database tables. Columns 4 
and 5 list the right ascension and declination of each object (J2000). The formal
uncertainties in the coordinates are 0.25 arcsec in RA and 0.20 arcsec in
declination.  Column 6 gives the B magnitude, while column 7 lists the 
B$-$V color.  For brighter objects, the magnitude estimates typically have 
uncertainties of 0.05 mag, increasing to $\sim$0.10 mag at B = 20.
Paper I includes a complete discussion of the precision of both the astrometry 
and photometry of the KISS objects.  An estimate of the redshift of each
galaxy, based on its objective-prism spectrum, in given in column 8.
This estimate assumes that the emission line seen in the objective-prism 
spectrum is H$\alpha$.  Follow-up spectra for $>$1200 ELG candidates from the 
two red survey lists (KR1 and the current list) show that this assumption is correct 
in the vast majority of cases.  Only 35 ELGs with follow-up spectra (2.8\%) are high 
redshift objects where a different line (typically [\ion{O}{3}] and/or H$\beta$) appears 
in the objective-prism spectrum.  Nine objects in the table have negative redshifts,
although only two of these are more than 1$\sigma$ below 0.0000.  We have obtained 
follow-up spectra for four of the nine: three are high redshift objects where the detected 
line is not H$\alpha$, and one is a low-z star-forming galaxy with an actual redshift of 
0.0006.   The formal uncertainty in the redshift estimates is $\sigma_z$ = 0.0028 (see Section 
4.1.3).  Columns 9 and 10 list the emission-line flux (in units of 10$^{-16}$ erg/s/cm$^2$) 
and equivalent width (in \AA) measured from the objective-prism spectra.
The calibration of the fluxes is discussed in section 4.1.2.  These quantities should be 
taken as being representative estimates only.  A simple estimate of the reliability of each 
source, the quality flag (QFLAG), is given in column 11.  This quantity, assigned during the 
line measurement step of the data processing, is given the value of 1 for high quality sources, 
2 for lower quality but still reliable objects, and 3 for somewhat less reliable sources.  Column 
12 gives alternate identifications for KISS ELGs which have been cataloged previously.  This 
is not an exhaustive cross-referencing, but focuses on previous objective-prism surveys which 
overlap part or all of the current survey area: Markarian (1967)  and Case (Pesch \& Sanduleak 
1983).   Also included are objects in common with the {\it Uppsala General Catalogue of 
Galaxies} (UGC, Nilson 1973).

A total of 1029 ELG candidates are included in this second list of H$\alpha$-selected 
KISS galaxies.  The total area covered by the second red survey strip is 65.8 deg$^2$, 
meaning that there are 15.6 KISS ELGs per square degree. For the first and second red 
lists combined, the surface density is 16.9 galaxies deg$^{-2}$. This compares to the 
surface density of 0.1 galaxies deg$^{-2}$ from the Markarian survey, and 0.56 
galaxies per deg$^2$ from the UCM survey; the present survey is much deeper!  
The second red survey strip has a somewhat lower surface density than the first
red survey strip. This could be an effect of the Bo\"otes void, which is included 
in the area covered by the second survey strip.

Of the 1029 objects cataloged, 630 were assigned quality values of QFLAG = 1 (61.2\%), 318 
have QFLAG = 2 (30.9\%), and 81 have QFLAG = 3 (7.9\%). Based on our follow-up spectra
to date, 98.6\% (206 of 209) of the sources with QFLAG = 1 are {\it bona fide}
emission-line galaxies, compared to 81.5\% (66 of 81) with QFLAG = 2 and 64.7\% (11 of 17)
with QFLAG = 3.  Overall, 92.2\% of the objects with follow-up spectra are {\it bona fide} 
ELGs.  The properties of the KISS galaxy sample are described in the next section.

\begin{figure*}[htp]
\vskip -0.5in
\epsfxsize=6.5in
\hskip 0.5in
\epsffile{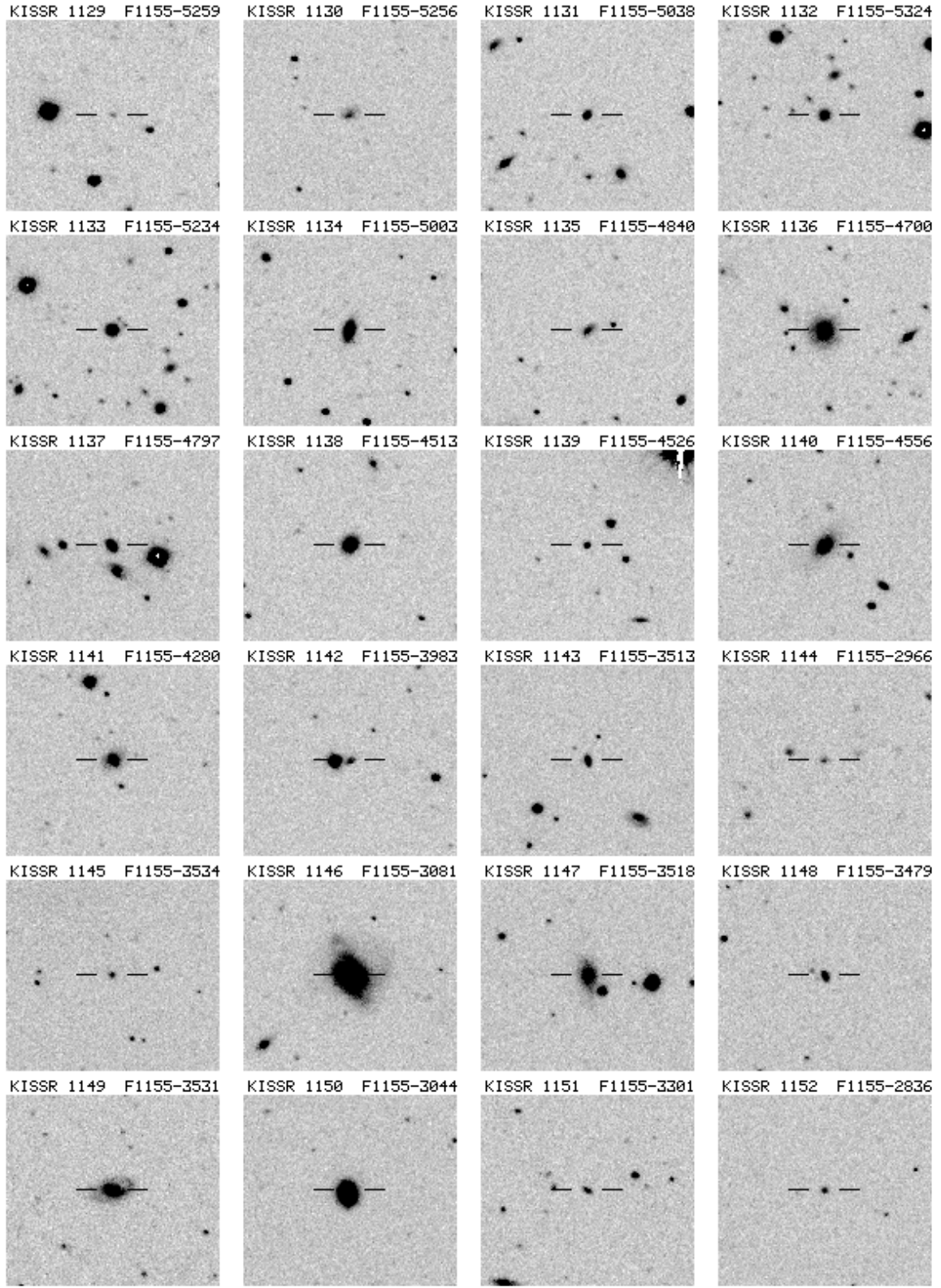}
\figcaption[find1.eps]{Example of finder charts for the KISS ELG candidates.
Each image is 3.2 $\times$ 2.9 arcmin, with N up, E left.  These finders are 
created from a composite of the B- and V-band direct images obtained as part 
of the survey.  In all cases the ELG candidate is located in the center of 
the image section displayed, and is indicated by the tick marks.\label{fig:find1}}
\end{figure*}

\begin{figure*}[htp]
\vskip -0.5in
\epsfxsize=6.5in
\hskip 0.5in
\epsffile{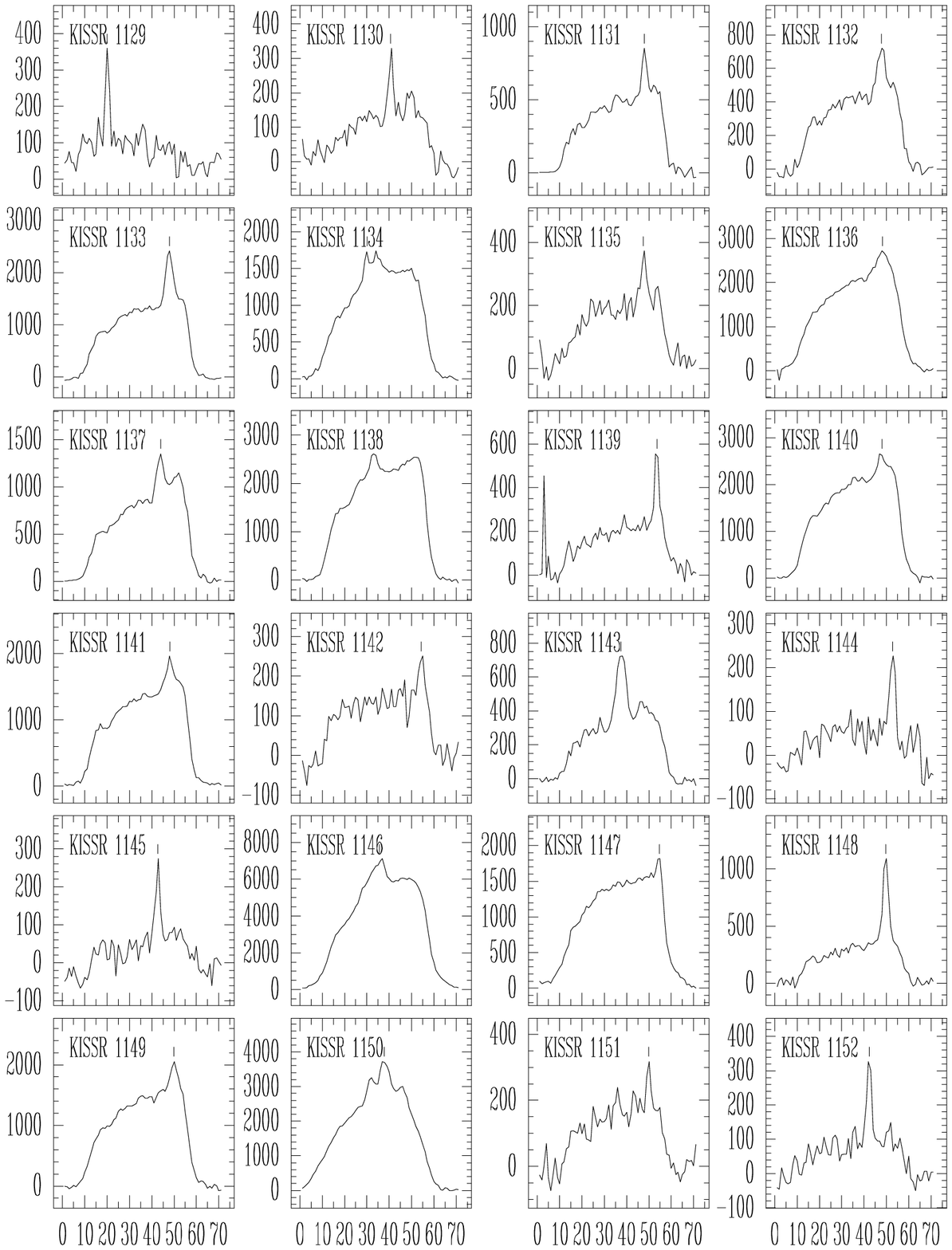}
\figcaption[spec1.eps]{Plots of the objective-prism spectra for 24 KISS ELG
candidates.  The spectral information displayed represents the extracted
spectra present in the KISS database tables.  The location of the putative 
emission line is indicated.\label{fig:spec1}}
\end{figure*}

Figure~\ref{fig:find1} shows an example of the finder charts for the KISS ELGs.  
These are generated from the direct images obtained as part of the survey, and 
represent a composite of the B- and V-band images.  Figure~\ref{fig:spec1} 
displays the extracted spectra derived from the objective-prism images for the 
first 24 ELGs in Table 1.  Finder charts and spectral plots for all 1029 objects 
in the current survey list are available in the electronic version of this paper. 

%\placefigure{fig:find1}

%\placefigure{fig:spec1}

A supplementary table containing an additional 291 ELG candidates is included 
in the appendix of this paper (Table \ref{table:tab3}).  These galaxies are 
considered to be lower probability candidates, having emission lines with 
strengths between 4$\sigma$ and 5$\sigma$. These additional galaxies do not constitute
a statistically complete sample, and should therefore be used with caution. However,
there are likely many interesting objects contained in this supplementary list.
Hence, following the precedent established in KR1, we list these objects in order
to give a full accounting of the ELGs in the area surveyed.

%************************************************************************

\section{Properties of the KISS ELGs}

The survey method employed by KISS makes it possible to obtain a general  
picture of the survey constituents, even in the absence of follow-up observations.
A large amount of information is available for each ELG candidate from the 
survey data themselves.  Accurate B and V photometry, astrometry, and morphological 
data can be derived from the direct images, and the digital objective-prism spectra 
allow us to measure line strengths and positions. However, due to the low resolution 
and limited spectral range of the objective-prism spectra, it is not possible to 
distinguish between different types of galactic activity based on the survey data 
alone. Follow-up spectroscopy is necessary to develop a more complete understanding 
of the nature of each ELG.  In the following section, we use the survey data to 
investigate the key properties of the second H$\alpha$-selected KISS ELG sample.

\subsection{Observed Properties}

%\placefigure{fig:appmag}
\begin{figure*}[htp]
\vskip -0.4in
\epsfxsize=5.0in
\hskip 1.0in
\epsffile{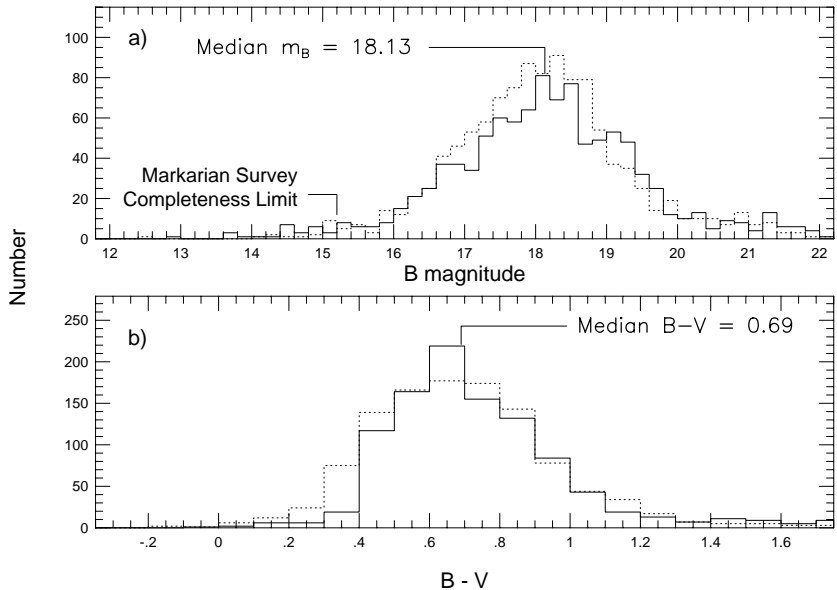}
\figcaption[fig3.eps]{(a) Distribution of B-band apparent magnitudes for the
1029 ELG candidates in the second H$\alpha$-selected KISS survey list.  
The median brightness in the KISS sample is B = 18.13, with 8\% of the 
galaxies having B $>$ 20.  Also indicated, for comparison, is the completeness 
limit of the Markarian survey. 
(b) Histogram of the B$-$V colors for the 1029 ELG candidates.  
The median color of 0.69 is indicated. 
The dotted lines show the magnitude and color distributions of the ELG 
candidates in the first KISS survey list.
\label{fig:appmag}}
\end{figure*}

\subsubsection{Magnitude \& Color Distributions}

Figure~\ref{fig:appmag}a plots the B-band apparent magnitude distribution for 
the 1029 KISS ELGs in the second red survey list, while Figure~\ref{fig:appmag}b
displays the distribution of B$-$V colors. The median values of both apparent 
magnitude and color are indicated. For comparison, the histogram for the 
first red survey list is shown in this and the following figures as a dotted line.

The median apparent B magnitude of the second survey list is 18.13. This value is very close 
to that of the first red survey list, which has a median apparent magnitude of B = 18.08.  
This compares to median apparent magnitudes of B = 16.9 for the [\ion{O}{3}]-selected 
Michigan (UM) survey (Salzer \etal 1989) and B $\approx$ 16.1 for the H$\alpha$-selected 
UCM survey (P\'erez-Gonz\'alez \etal 2000).  
The completeness limit of the Markarian survey, B = 15.2 (Mazzarella \& Balzano 1986) is
indicated in Figure~\ref{fig:appmag}a. The KISS sample clearly probes substantially 
deeper than previous objective-prism surveys. 

The median color of the second red survey list is B$-$V = 0.69, which is nearly identical 
to the median color of the first red survey list, B$-$V = 0.67.
Both of these median values fall in the typical color range of an Sb galaxy (Roberts \& 
Haynes 1994). The color distribution appears to be similar to that of the UCM galaxies
(P\'erez-Gonz\'alez \etal 2000), which have a mean B$-$r color of 0.71. This mean value 
is comparable to the mean color of an Sbc galaxy (Fukugita \etal 1995). The KISS
red survey and the UCM survey galaxies are both significantly redder than the KISS blue 
survey (median color of B$-$V = 0.50; KB1), or the UM survey (median color of 0.54; Salzer 
\etal 1989). The reason for this difference can be found in the selection methods of the 
surveys. The UM and KB1 surveys are both [\ion{O}{3}]-selected, hence they preferentially 
detected ELGs with lower luminosities and with low amounts of reddening. In contrast, 
H$\alpha$-selected surveys tend to detect a larger proportion of more luminous starburst 
galaxies and AGN, and they are significantly less biased against heavily reddened ELGs.  
In addition to these intrinsically red and/or higher reddening galaxies, the H$\alpha$-selected 
KISS sample also contains a large number of dwarf galaxies (see Section 4.2.1). Hence it 
exhibits a color range that more closely matches that of the overall galaxian population.

The color distribution shows an extended tail of very red ELGs. A total of 62 KISS galaxies 
in the current list possess B$-$V colors $>$ 1.2 (i.e., redder than any unreddened
stellar population typical of a normal galaxy).  
The median apparent magnitude of these red objects is B = 20.7. At such faint magnitudes 
the uncertainties in the photometry are substantial (typical errors in the B$-$V color are 
0.2 to 0.3 magnitude).
Spectroscopic follow-up observations have been obtained for nine of these very red ELG 
candidates, and only three have been confirmed as ELGs. The remaining six are stars or faint 
galaxies with no emission lines. These results indicate that many of these faint, red 
candidates are probably false detections; the proportion of spurious sources is known to 
increase at fainter magnitudes.  This is reflected in the fact that 89\% of these red objects
(55 of 62) were assigned quality codes of 2 or 3.  However, all three of the confirmed red
ELGs mentioned above are Seyfert galaxies, which suggests that follow-up spectra will 
reveal some interesting objects in this population of very red KISS objects.

\subsubsection{Line Strength Distributions and Survey Completeness}

%\placefigure{fig:ew}
\begin{figure*}[htp]
\vskip -0.4in
\epsfxsize=5.0in
\hskip 1.0in
\epsffile{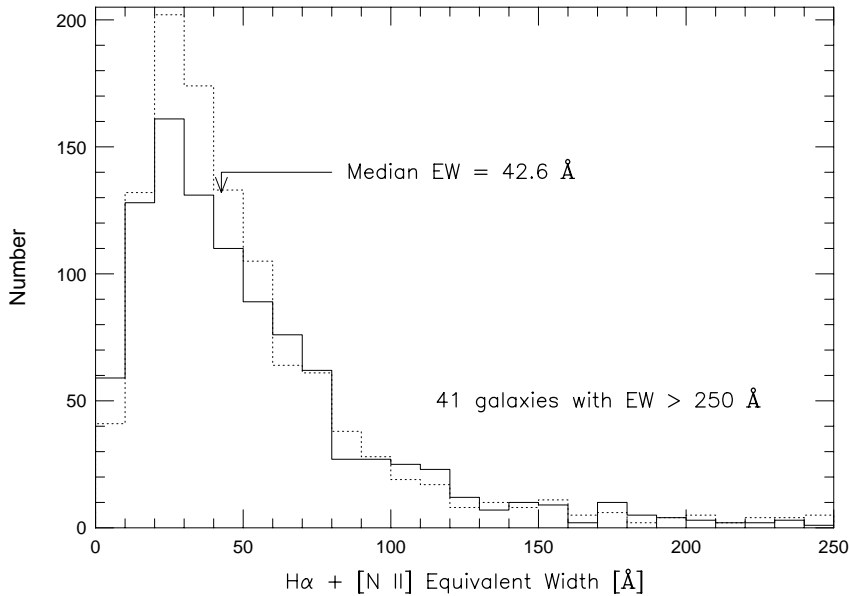}
\figcaption[fig4.eps]{Distribution of measured H$\alpha$ + [\ion{N}{2}]
equivalent widths for the KISS ELGs.  The median value of 42.6 \AA\ is 
indicated.  The measurement of equivalent widths from objective-prism 
spectra tends to yield underestimates of the true equivalent widths, so
these values should be taken only as estimates.  The survey appears to 
detect most sources with EW(H$\alpha$+[\ion{N}{2}]) $>$ 20--30 \AA.
The dotted line shows the equivalent width distribution of the ELG candidates 
in the first KISS survey list.
\label{fig:ew}}
\end{figure*}

The digital nature of the survey data, along with our
KISS software, allow us to measure the position and strength of the
emission lines that are seen in the objective-prism spectra of the KISS
candidates.  We are able to measure both the emission line flux and
equivalent width for each source, which allows us to utilize the survey 
data themselves to investigate issues like the star-formation rate density
in the local universe.  In addition, we are able to assess the completeness
of our ELG sample directly from the survey data.  Since KISS is a
line-selected survey, its completeness limit must be defined in terms of line 
strengths, not continuum apparent magnitudes (Salzer 1989).  Gronwall 
\etal (2004) describes our method for determining the completeness of
the survey and derives an estimate of the local star-formation rate density,
using the KR1 sample.

For the reasons discussed in KR1, we do not expect our measured line
strengths to be as precise as those obtained from slit spectra.   In
particular, equivalent widths measured from extended sources will tend 
to be underestimated due to the slitless nature of the objective-prism
spectra.  Since continuum light from many portions of the galaxy can be
dispersed by the prism to the location of the emission line in the
objective-prism spectrum, this effect can be large, leading to underestimates 
of factors of two or more in the survey equivalent widths!  Wegner \etal (2003)
presents a thorough comparison of the equivalent widths measured from both
the survey data and slit spectra for over three hundred KISS ELGs.
On the other hand, we expect that our measured line {\it fluxes} 
should be more representative of the total emission from our galaxies
than will fluxes measured from slit spectra, especially for extended
objects.  This is because we sum over 5 pixels (7.3 arcsec) when we 
extract the objective-prism spectra, compared to a typical slit width
of 1 -- 2 arcsec for our follow-up spectra.  Furthermore, the field-to-field
flux calibration employed (Paper I) ensures that the entire survey is
on precisely the same relative flux scale.  Thus,  the KISS line fluxes 
should represent fairly accurate estimates of the total H$\alpha$ + 
[\ion{N}{2}] emission.

We display the distribution of equivalent widths seen in our survey galaxies 
in Figure~\ref{fig:ew}.   Since we are without the benefit of follow-up spectra
for the majority of our candidates, we must assume that the line we are 
measuring is H$\alpha$ + [\ion{N}{2}].  Based on our follow-up spectra to 
date (see Section 3.2), this appears to be a reasonable assumption.
The [\ion{N}{2}]$\lambda\lambda$6584,6548 lines are always blended with 
H$\alpha$ at our spectral resolution; there is no way to determine
the contribution from the individual lines using our survey data.  However,
the [\ion{S}{2}]$\lambda\lambda$6731,6717 doublet is well resolved from the 
H$\alpha$ + [\ion{N}{2}] complex, and is often seen in the objective prism 
spectra of strong-lined objects.  The median equivalent width found from the 
current sample of ELGs is 42.6 \AA.   This agrees closely with the median 
equivalent width of the first survey list, which is 41.0 \AA.  The vast majority 
of the galaxies have equivalent widths less than 100 \AA.  The distribution of
EWs peaks in the 20--30 \AA\ bin, suggesting that KISS is fairly complete
for objects with EWs above $\sim$30 \AA.

%\placefigure{fig:lfluxcal}
\begin{figure*}[htp]
\vskip -0.4in
\epsfxsize=5.0in
\hskip 1.0in
\epsffile{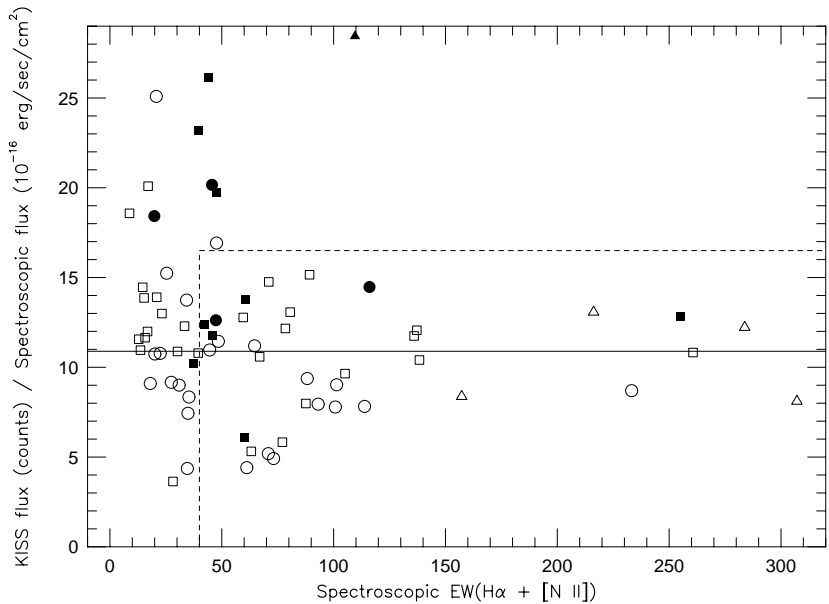}
\figcaption[fig5.eps]{Plot of the ratio of objective-prism flux (in counts) 
to spectroscopic flux {\it versus} H$\alpha$~+~[\ion{N}{2}] equivalent width measured 
from the follow-up spectra.  Data from six different observing runs on three 
different telescopes are plotted: open square = MDM 2.4-m in May, 2000, 
open circle = KPNO 2.1-m in July 2000, open triangle =
Lick 3.0-m in April, 2001, filled square = MDM 2.4-m in April, 2002,
filled triangle = Lick 3.0-m in April, 2002, and filled circle = MDM 2.4-m
in April, 2003.  The solid line indicates the median ratio.  The dashed lines show the 
criteria we applied to select the calibration sample. Four galaxies with 
flux ratios $>$~30 fall above the diagram, and one galaxy with 
EW $>$~320~\AA\ lies off the diagram to the right.
\label{fig:lfluxcal}}
\end{figure*}

The calibration of the flux scale is a two-step process.  First, the objective-prism 
spectra for each field are corrected for throughput variations and atmospheric 
extinction.  This places all line fluxes on the same {\it relative} flux scale.  Then, 
the fluxes are calibrated using information obtained from our follow-up spectra.
At the time of this writing, we have obtained slit spectra for 307 KISS candidates
from the sample of 1029 presented in this paper.  From these, we select galaxies 
that are starforming (i.e., not AGN), are of high quality, and had been observed 
with a long-slit spectrograph under photometric conditions.  Galaxies observed 
through optical fibers were not used, since such spectra are
notoriously difficult to use for accurate spectrophotometry.  This limited our 
calibration sample to 75 galaxies.  All data used were obtained using slit
widths of either 1.7 or 2.0 arcsec.  These wide slits guarantee that we recorded 
most of the emission-line flux from the sources, as long as the emission regions 
were not spatially extended.  Since the fluxes measured from the 
objective-prism spectra are a combination of the  H$\alpha$ and [\ion{N}{2}]
lines, we used the fluxes from our slit spectra for the sum of these three
lines.  Figure~\ref{fig:lfluxcal} shows a plot of the ratio of objective-prism
flux (in counts) to spectroscopic flux {\it versus} the equivalent width 
measured from the follow-up spectra.  Data from six different observing runs 
on three different telescopes are plotted, and there are no obvious systematic 
trends in this ratio with either equivalent width or observing run.  There are, 
however, a number of galaxies with large ratios, particularly at low equivalent 
width.  A visual inspection of the survey images showed that these galaxies 
are of large angular extent and exhibit extended line emission. The high flux 
ratios for these objects indicate that our long-slit measurements do not include 
all of the H$\alpha$ emission from these sources.  Thus, we restricted our 
analysis to those galaxies with an objective-prism-to-spectroscopic flux ratio 
of less than 16.5 and an equivalent width greater than 40 \AA, leaving us with 
a calibration sample of 38 galaxies.  
These galaxies all possess emission regions that are essentially point sources.
The median flux ratio of this sample was 10.89; the mean was 10.25 with a standard 
deviation of 2.96 and an error in the mean of 0.48.  We have adopted as our calibration 
value the reciprocal of the median value, or 0.0918 $\times 10^{-16}$ ergs/sec/cm$^{-2}$ 
per count. 

\begin{figure*}[htp]
\vskip -0.4in
\epsfxsize=5.0in
\hskip 1.0in
\epsffile{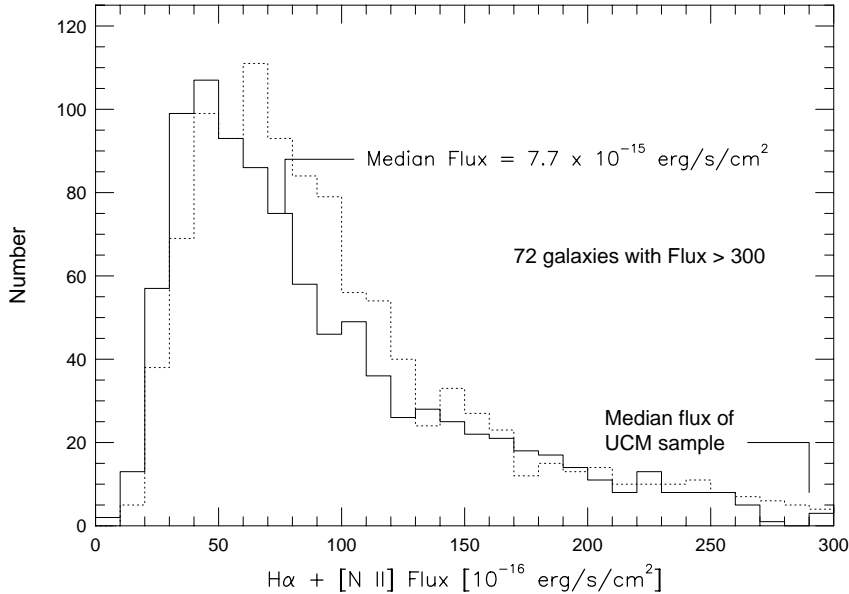}
\figcaption[fig6.eps]{Distribution of H$\alpha$ + [\ion{N}{2}] line fluxes
for the 1029 KISS ELGs included in the current survey list.  The median flux
level of both the KISS and UCM samples is indicated.
The dotted line shows the line flux distribution of the ELG candidates 
in the first KISS survey list. 
\label{fig:lflux}}
\end{figure*}

This calibration value is applied to our measured objective-prism line fluxes to
convert their instrumental fluxes (in counts) to calibrated fluxes (in erg/s/cm$^2$).
The distribution of  observed H$\alpha$+[\ion{N}{2}] line flux values for the 1029 
KISS ELGs is displayed in Figure~\ref{fig:lflux}.   The median value of 7.7 $\times$ 
10$^{-15}$ erg/s/cm$^2$ is somewhat lower than the first survey list median flux of 
8.7 $\times$ 10$^{-15}$ erg/s/cm$^2$.  This suggests that the data used for the current 
survey list are slightly more sensitive than those used for KR1, which is likely due to
the improved pixel scale and hence better spectral resolution achieved with the new
CCD.  For comparison, the median flux from the UCM sample is 
2.9 $\times$ 10$^{-14}$ erg/s/cm$^2$ (based on follow-up spectra of Gallego \etal 
1996).  A galaxy with this latter flux level would fall at the extreme right edge of the 
figure, which illustrates the increased depth of KISS relative to the UCM survey
in terms of line flux.

%\placefigure{fig:lflux}

\begin{figure*}[htp]
\vskip -0.4in
\epsfxsize=5.0in
\hskip 1.0in
\epsffile{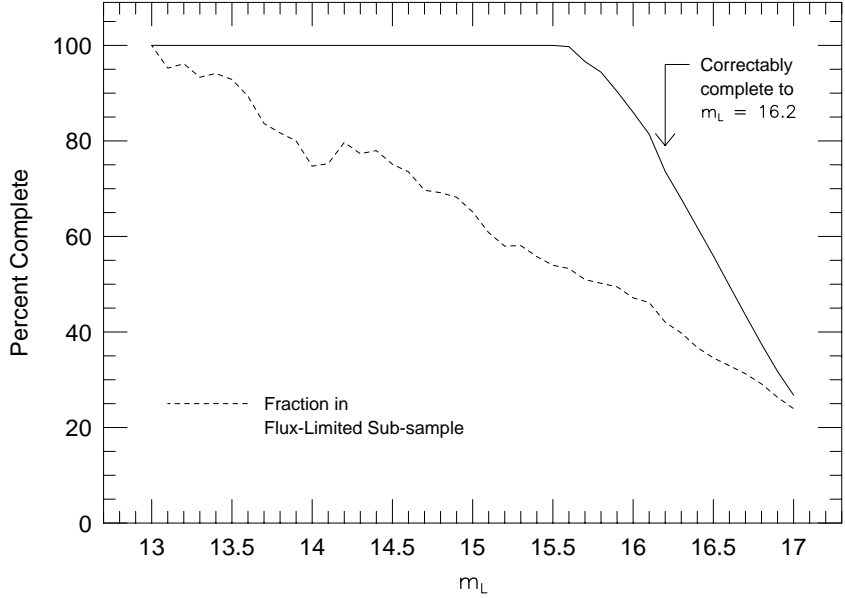}
\figcaption[complete_43red.eps]{Plot of the completeness percentage as a function of 
m$_L$ for the current sample (solid line).  The catalog is 100\% complete to
m$_L$ = 15.5, and is ``correctably complete" to m$_L$ = 16.2.  The dashed line
shows the fraction of the sample contained in the flux-limited sub-sample as
a function of m$_L$.  At the completeness limit, roughly half of the KISS 
ELGs are in the flux-limited portion of the sample.
\label{fig:complete}}
\end{figure*}

The calibrated line fluxes are used to determine the completeness limit of the
survey.    Using the procedure described in Gronwall \etal (2004), we convert these
fluxes into pseudo-magnitudes --  the line magnitude m$_L$.  We then apply a V/V$_{max}$
analysis (e.g., Schmidt 1968, Huchra \& Sargent 1973) to the full sample of 1029 
galaxies.  The results are shown in 
Table~\ref{table:comptab}.  The contents of the table are as follows: Column (1)
lists m$_{comp}$, the value of m$_L$ for which $\langle V/V_{max}\rangle$
is being computed.  Column (2) lists the total number of ELGs brighter 
than that m$_L$ level, while columns (3) and (4) gives the numbers of objects
in the volume-limited and flux-limited subsamples, respectively.  Note that
some objects may start out in the flux-limited sample at brighter values of
m$_{comp}$, then move into the volume-limited sample at fainter values of
m$_{comp}$.  Column (5) lists the mean V/V$_{max}$ for the flux-limited 
subsample.  Column (6) shows the number of galaxies that need to be added 
to the sample at each m$_{comp}$ level to maintain $\langle V/V_{max}\rangle 
= 0.5$, and column (7) lists the cumulative number of galaxies added at all 
magnitudes brighter than the given magnitude level to maintain
$\langle V/V_{max}\rangle$.  Column (8) shows the percentage of objects that
are in the flux-limited subsample, which decreases continuously as m$_L$ becomes
fainter.  Column (9) lists the completeness percentage of the flux-limited 
subsample as a function of m$_L$.   These latter two quantities are plotted in 
Figure~\ref{fig:complete}.

%\placefigure{fig:complete}

The interpretation of the V/V$_{max}$ test follows
exactly the discussion found in Gronwall \etal (2004) for the KR1 sample, and hence 
will not be repeated here.  Rather, we will simply summarize the main results.
First, we note that because of the redshift limit imposed by the filter used for the 
survey, objects in the sample can be either line-flux-limited or volume-limited objects,
depending on the strength of their H$\alpha$+[\ion{N}{2}] emission and their actual 
redshift.  As the limiting line flux (parameterized by m$_L$) decreases, a given 
object can actually switch from being in the flux-limited category to the volume-limited
category.  For faint limiting line fluxes (fainter m$_L$) the majority of the KISS
ELGs are in the volume-limited subsample.  This is illustrated by the dashed line in 
Figure~\ref{fig:complete}.  Second,  we see that the KISS sample is 100\% complete down 
to m$_L$ = 15.5, which is nearly identical to the result for the KR1 sample (Gronwall 
\etal 2004).  This completeness limit includes 630 KISS ELGs, or 61.2\% of the sample.  
One can construct a ``correctably complete" sample by extending the line-flux limit 
down to even lower values.  For example, at m$_L$ = 16.2, the sample is still 73.6\% 
complete, but now includes 89.7\% of the sample.

%\placetable{table:comptab}

\subsubsection{Redshift Distributions}

\begin{figure*}[htp]
\vskip -0.4in
\epsfxsize=5.0in
\hskip 1.0in
\epsffile{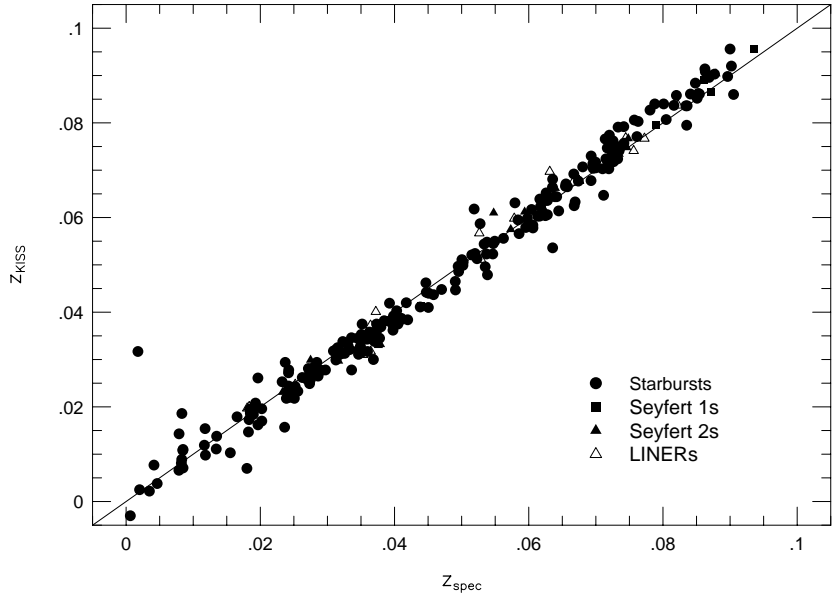}
\figcaption[zcomp.eps]{Comparison between objective-prism redshifts (z$_{KISS}$) 
obtained from our survey data and slit-spectra redshifts (z$_{spec}$) obtained from
follow-up spectra.  The solid line denotes z$_{KISS}$ = z$_{spec}$.   The objective-prism
redshifts provide reasonable estimates of the true redshifts over the full range covered
by the survey.  The formal uncertainty in z$_{KISS}$ is 0.0028 (840 \kms).
\label{fig:zcompare}}
\end{figure*}

Another parameter that we derive from the objective-prism spectra is the 
redshift of each object.  Following the example from Paper I, we can compare the
objective-prism redshifts obtained from the survey data with redshifts derived from
slit spectra for those objects for which follow-up spectra have been obtained. 
This allows us to assess the uncertainty associated with the objective-prism
redshifts.  Figure~\ref{fig:zcompare} plots z$_{KISS}$ (objective-prism redshift)
against z$_{spec}$ (follow-up spectra redshifts), and in general shows excellent
agreement between the two measurements.  A few objects deviate substantially from
the equality line.  We checked several such objects in the original survey data and 
discovered that, in all cases, the emission region is offset from the center of the galaxy 
in a north-south direction.  Because the dispersion of the objective-prism spectra 
is in the N-S direction, and because of the slitless nature of the spectra and the 
manner in which they are extracted (see Paper I), objects with emission regions
that are off-center will yield unreliable redshift estimates from their objective-prism
spectra.  This is the case for only a small minority of KISS objects.  The RMS scatter 
of z$_{KISS}$ about the equality line is 0.0036 (1080 \kms).  However, if the three
most deviant objects are removed from the diagram, the RMS scatter is reduced to
0.0028 (840 \kms).  We adopt this latter value as our redshift precision.  It is
precisely the same value found for the KR1 sample (Paper I).

%\placefigure{fig:zcompare}

%\placefigure{fig:zhist}
\begin{figure*}[htp]
\vskip -0.4in
\epsfxsize=5.0in
\hskip 1.0in
\epsffile{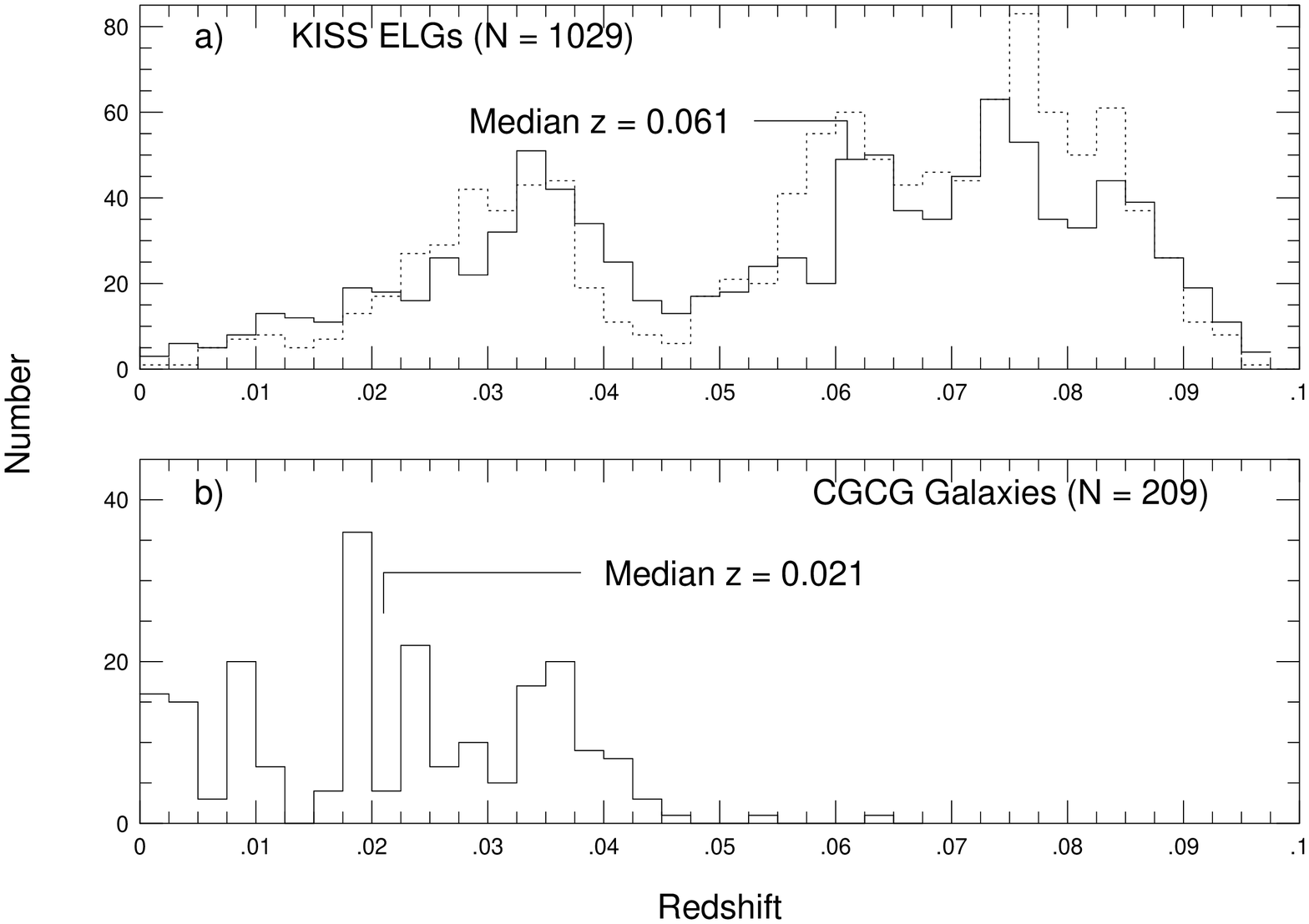}
\figcaption[fig7.eps]{Histograms showing the distribution of redshift for $(a)$
the 1029 H$\alpha$-selected KISS ELGs and $(b)$ the 209 ``normal" galaxies from the
CGCG that are located in the same area of the sky. The dotted line in the top panel 
shows the redshift distribution of the ELG candidates in the first KISS survey list.
The median redshift is indicated in both plots.  Note that the number of KISS ELGs 
continues to rise up to the cut-off of the filter used for the survey, indicating that 
the survey is volume-limited for the more luminous galaxies.  The deficit of ELGs 
between z = 0.04 and 0.06 is due to the Bo\"otes void. 
\label{fig:zhist}}
\end{figure*}

For objects in the KR1 survey list it was necessary to apply a correction to objective-prism 
redshifts above $z = 0.07$ (the details are given in Paper I).  Above this redshift, the value
of z$_{KISS}$ systematically underestimated the true redshift because the H$\alpha$ line
begins to redshift out of the survey bandpass.   However, the objects plotted in  
Figure~\ref{fig:zcompare} do not display this offset: the objective-prism redshifts agree well 
with redshifts determined from follow-up spectroscopy out to the $z \sim 0.1$ cutoff imposed 
by the survey filter.  The reason for this difference is probably the better pixel scale of the 
CCD used for the second red survey.  We do not apply any corrections to the objective-prism
redshifts listed in Table~\ref{table:tab2}

Figure~\ref{fig:zhist} illustrates the distribution of the objective-prism redshifts.
The upper panel shows the redshift distribution of the KISS ELGs from the second 
red survey list as a solid line, and the distribution of the first red survey list 
is indicated by a dotted line. 
The lower panel shows the redshift distribution for a comparison sample of galaxies 
from Zwicky \etal (1961--1968; hereafter CGCG). The redshifts for the 209 CGCG galaxies 
are taken from Falco \etal (1999); this is essentially the portion of the CfA2 
redshift survey that covers the same area on the sky as the KISS ELGs. The Falco 
\etal redshift catalog sample is complete to m$_B$ = 15.5. 
Because the surface density of the CGCG catalog is fairly low, we included objects 
lying in a four-degree-wide declination strip, rather than the 1.6 degree covered 
by KISS. We use these CGCG galaxies as our comparison sample in the following section as well.

From Figure~\ref{fig:zhist} it is immediately apparent that KISS detects large 
numbers of galaxies to much higher redshifts than what is found in the CGCG sample. 
The median redshift for the KISS sample is nearly three times greater than that 
of the magnitude-limited CGCG galaxies.
The Bo\"otes void is the reason for the large drop in the number of galaxies 
between z = 0.04 and 0.06 (see also Figure~\ref{fig:cone2}).
Only a handful of CGCG galaxies are found beyond this void.  In contrast,
the number of KISS ELGs increases again beyond the Bo\"otes void, out to the 
redshift limit imposed by the survey filter.  The implications
of this are (1) that the survey technique employed by KISS would be sensitive to
galaxies at distances well beyond the distance limit imposed by the filter, if
the filter was either not used or was replaced with one that extended to
redder wavelengths, and (2) for the high-luminosity portion of the ELG luminosity 
function, the KISS sample is effectively volume-limited rather than flux-limited. 

Comparison of the velocity histograms of the two KISS lists (Figure~\ref{fig:zhist}a)
shows that they are very similar.  In particular, the KR1 histogram shows a
depression in the number of galaxies at velocities comparable to the the Bo\"otes
void (12,000 -- 18,000 \kms).  While this may appear strange at first sight, since
the two survey strips are separated by 14$\arcdeg$ in declination, the underdensity 
associated with the Bo\"otes void is in fact quite large.  For example, the
orignal paper announcing the discovery of the void (Kirshner \etal 1981) was based
on redshift survey data separated by 43$\arcdeg$ in declination (26$\arcdeg$ to 69$\arcdeg$).
Hence, the similarities in the two histograms is most likely indicating the linkage in
the large-scale spatial distribution of the galaxies in these two widely separated 
fields.

\subsection{Derived Properties}

\subsubsection{Luminosity Distribution}

%\placefigure{fig:absmag}
\begin{figure*}[htp]
\vskip -0.4in
\epsfxsize=5.0in
\hskip 1.0in
\epsffile{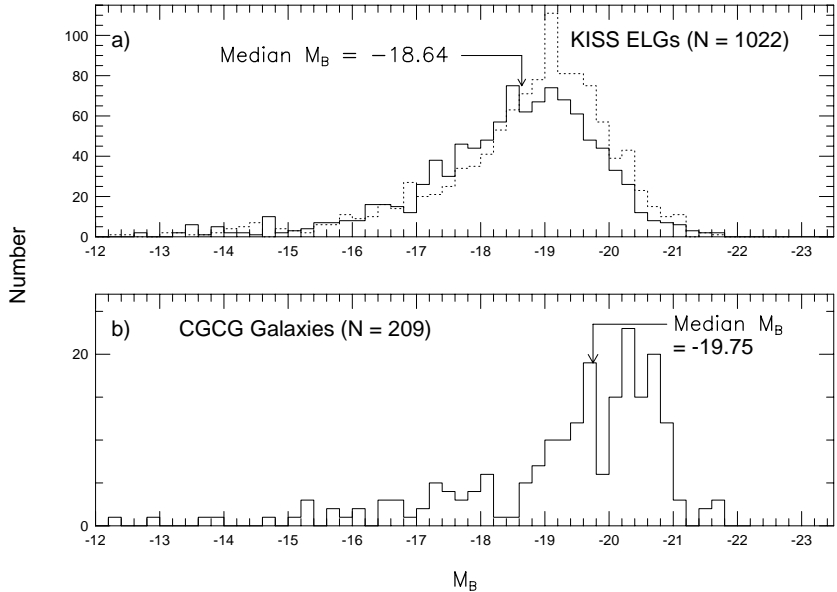}
\figcaption[fig8.eps]{Histograms showing the distribution of blue absolute 
magnitude for $(a)$ the 1029 H$\alpha$-selected KISS ELGs and $(b)$ the 209 
``normal" galaxies from the CGCG that are located in the same area of the sky.  
The dotted line shows the luminosity distribution of the ELG candidates 
in the first KISS survey list. The median luminosity of each sample is indicated.  
The KISS ELG sample is made up of predominantly intermediate- and 
lower-luminosity galaxies, making this line-selected sample particularly 
powerful for studying dwarf galaxies. 
\label{fig:absmag}}
\end{figure*}

Using the redshifts and apparent magnitudes listed in Table \ref{table:tab2} and 
assuming a value for the Hubble Constant of H$_o$ = 75 km/s/Mpc, we compute 
absolute magnitudes for the second red survey list.
Corrections for Galactic absorption (A$_B$) have been applied by averaging the 
absorption values for all KISS emission-line galaxies in each survey field. 
The extinction value for each ELG position was extracted from the maps of dust 
infrared emission constructed by Schlegel, Finkbeiner, \& Davis (1998).  
Since the majority of the survey strip is at high Galactic
latitude, this correction is typically small: 55 of the 62 fields (89\%) have 
A$_B$ $<$ 0.10, and all fields have A$_B$ $<$ 0.13.  The maximum correction
of A$_B$ = 0.123 occurs in one of the easternmost survey fields (F1532).
Note also that the large redshift uncertainties associated with the 
objective-prism spectra (Section 4.1.3) translate into significant uncertainties
in our computed absolute magnitudes.  For example, an object with the median
redshift (z = 0.061) will have an uncertainty in its M$_B$ of 0.10 magnitude
due to the redshift uncertainty alone.  The error is much larger for nearby
objects ($\sim$0.4 magnitude for a galaxy with z = 0.01).  These uncertainties 
should be kept in mind when interpretting Figure~\ref{fig:absmag}.

We compare the luminosities of the KISS ELGs with those of the CGCG galaxies
located in the same area of the sky in Figure~\ref{fig:absmag}. 
The median blue absolute magnitude of the KISS ELGs is $-$18.64, which is roughly
one magnitude fainter than M$^*$, the ``characteristic luminosity" parameter
of the Schechter (1976) luminosity function.  As seen in the lower portion of the
figure, the majority of the CGCG galaxies are more luminous than the KISS ELGs.  
The median absolute magnitude of the CGCG sample is $-$19.75 (i.e., very close 
to M$^*$). 

We note that in this survey region, both the KISS sample and the CGCG samples 
have median luminosities that are roughly one-third of a magnitude fainter than
those of the corresponding samples in the KR1 survey strip. This could be 
due to the fact that the second red survey strip runs through the center of the Bo\"otes 
void.  Since KR1 and the current survey list have comparable depths and similar
velocity cutoffs, the main difference between them would appear to be a deficiency
of galaxies at intermediate redshifts caused by the void.  This could explain the
differences in the median luminosities as due to the fact that the median and high
luminosity objects that would naturally be visible at the distance of the void
are ``missing" from the sample.  However, as noted above, KR1 has a similar depression
in its velocity histogram in roughly the same velocity range, so this explanation
may not be wholly correct.

%\placefigure{fig:cone1}
\begin{figure*}[htp]
\vskip -0.4in
\epsfxsize=5.7in
\hskip 0.5in
\epsffile{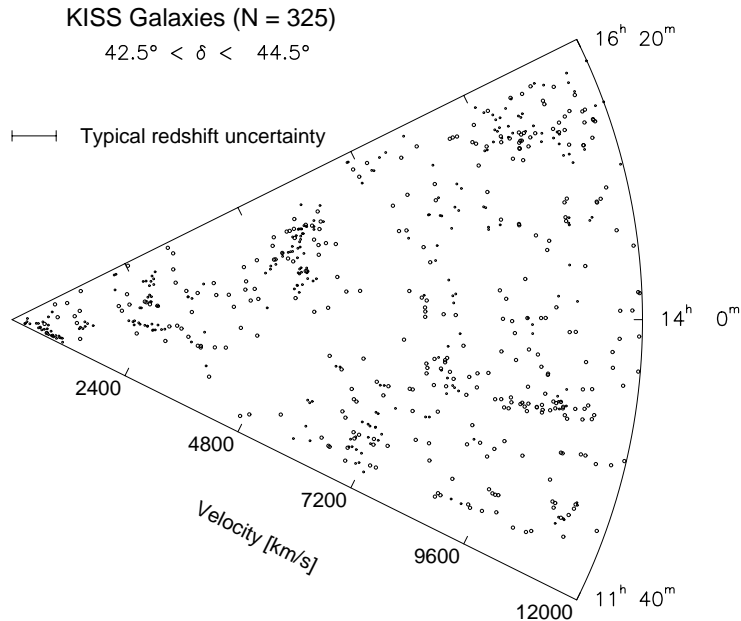}
\figcaption[cone1.eps]{This plot compares the spatial distribution of ELGs to 
that of ``normal" galaxies, out to velocities of 12,000 \kms. The ``normal" galaxies
are represented by CGCG galaxies from the declination range $41\arcdeg < \delta < 
46\arcdeg$. Only 12 CGCG galaxies are found at velocities $>$ 12,000 \kms.  The KISS 
galaxies are plotted as open circles, and the CGCG galaxies are overplotted as dots. 
At lower velocities the ELGs are seen to trace the large-scale structures defined by 
the CGCG galaxies.  However, they appear to be less tightly clustered, and a large 
number of ELGs fall in voids. 
\label{fig:cone1}}
\end{figure*}

Figure~\ref{fig:absmag} shows how the KISS sample is dominated
by intermediate- and low-luminosity galaxies, with a typical luminosity 
comparable to that of the Large Magellanic Cloud.  
A significant population of higher luminosity galaxies is also present, 
but the high-luminosity end of the distribution appears truncated when 
compared to the CGCG sample. This effect is due to the filter-induced redshift 
cut-off of the KISS sample, which causes the luminous end of the sample 
to become volume-limited. Hence, the KISS sample lacks the high
luminosity tail that is commonly seen in magnitude-limited samples.

Even though there are large numbers of lower luminosity ELGs present in
KISS, the proportion of dwarf star-forming systems is significantly lower
than in [\ion{O}{3}]-selected surveys like the UM survey and KB1.  The median
absolute magnitude for the UM ELGs is M$_B$ = $-$18.1 (Salzer \etal 1989),
while that of KB1 is $-$18.0.  Despite these differences in the median luminosities 
of the [\ion{O}{3}]-selected vs. the H$\alpha$-selected samples, one should not 
conclude that the KISS sample is deficient in dwarf galaxies relative to the blue 
surveys.   As demonstrated by the large overlap between the first blue survey list 
(KB1) and the first red survey list (KR1), the H$\alpha$ survey technique is just as 
efficient at finding dwarf ELGs as is the [\ion{O}{3}]-selection method.
However, H$\alpha$-selected samples can detect many more higher-luminosity 
galaxies which tend to be missed by the [\ion{O}{3}]-selected surveys.
More luminous galaxies with starbursts tend to have weaker [\ion{O}{3}] lines.
In addition, [\ion{O}{3}]-selected surveys tend to be biased against 
luminous galaxies due to the higher amounts of reddening present in these
more metal-rich systems. Since KISS selects by H$\alpha$, it does not
suffer from these biases. The KISS sample should thus be a more
representative catalog of AGN and star-forming galaxies.

\subsubsection{Spatial Distribution}

\begin{figure*}[htp]
\vskip -0.4in
\epsfxsize=5.7in
\hskip 0.5in
\epsffile{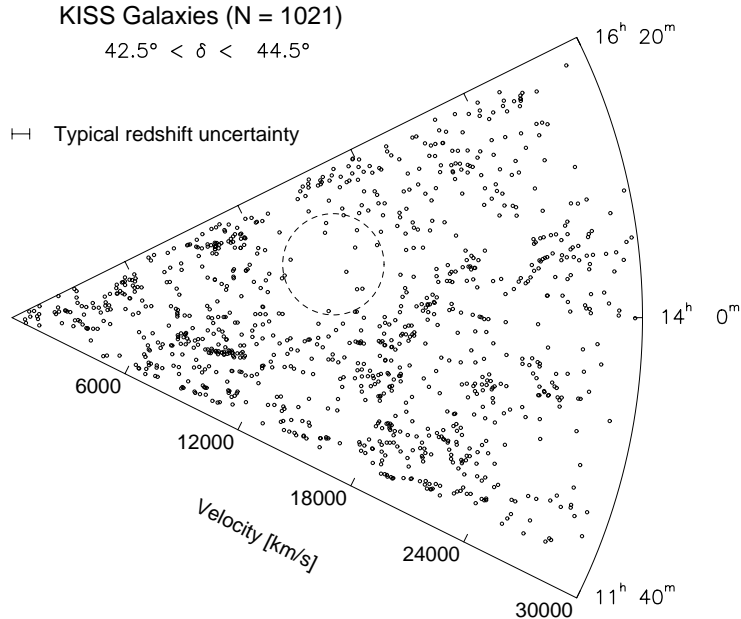}
\figcaption[cone2.eps]{The spatial distribution of KISS galaxies is shown out
to velocities of 30,000 \kms. The circle indicates the approximate position of 
the Bo\"otes void.  The numbers of ELGs remains high out to $\sim$27,000 \kms, 
the point where the survey filter cut-off begins to affect the sample. The KISS
catalog is sufficiently deep to adequately sample the far side of the Bo\"otes 
void. We also find 12 objects that appear to be located inside this void.
\label{fig:cone2}}
\end{figure*}

Because of the depth and good redshift coverage of the KISS ELGs, it is also 
relevant to consider the spatial distribution of the sample.  This is illustrated in 
Figures~\ref{fig:cone1} and \ref{fig:cone2}, which show cone diagrams for the 
KISS and CGCG galaxies.   The redshifts plotted for the KISS galaxies are those 
from Table \ref{table:tab2}.  The error bar shown in both figures illustrates the
size of the formal uncertainty in our objective-prism redshifts (840 \kms, see
Section 4.1.3).

Figure~\ref{fig:cone1} shows both the KISS and CGCG galaxies out to 12,000 \kms 
(z $\le$ 0.04). At redshifts beyond about 12,000 \kms, the CGCG sample thins out 
drastically, and no longer can be used to delineate large-scale structure.
The general impression obtained from examination of this figure is that the
KISS ELGs tend to fall along the same large-scale structures as those traced out
by the CGCG galaxies.  However, the ELGs appear to be less tightly confined to the 
structures than are the ``normal" galaxies.  Because of the limited precision of 
the KISS redshifts ($\pm$840 \kms), it is possible that the appearance of lower 
clustering is an artifact of the data (i.e., consider the size of the typical 
redshift error illustrated).  However, the large numbers of galaxies located well 
into some of the voids present within this volume of space suggest that the lower 
level of clustering is in fact real.  Previous studies of the spatial distribution of 
ELG samples have found similar results (Salzer 1989, Rosenberg \etal 1994, Pustil'nik 
\etal 1995, Popescu \etal 1997, Lee \etal 2000).  The lower level of clustering seen
in Figure~\ref{fig:cone1} is due primarily to lower-luminosity ELGs.  We are currently 
analyzing the relative clustering strengths of the ELGs and CGCG  galaxies, using 
more accurate redshifts from our follow-up spectra when available (Lee, Salzer \& 
Gronwall 2004).

Figure~\ref{fig:cone2} shows all KISS galaxies out to 30,000 \kms (z $\le$ 0.10).
The KISS galaxies are found in great numbers out to $\sim$27,000 \kms, beyond 
which the survey begins to be truncated by the filter.   The approximate position of 
the Bo\"otes void is indicated by the circle.  We have chosen a rather conservative
estimate for the radius of the void, r$_{void}$ = 2,400 \kms (most previous studies
have adopted r$_{void}$ = 3,000 \kms).  It is clear from this figure that the depth 
of the KISS survey is sufficient to adequately sample the far side of the void.  
Based on our objective-prism redshifts, we find 12 objects that appear to be located 
inside the Bo\"otes void (18 if we were to use r$_{void}$ = 3,000 \kms), only one of 
which was known previously to be a void-dwelling object.  The void galaxies have a 
median B magnitude of 18.25 (range 17.38 to 20.73) and a median absolute B magnitude 
of $-$18.56 (range $-$16.02 to $-$19.07).  Only one of the twelve galaxies has been
observed spectroscopically by our group at this time.  That object is KISSR 1943, 
which is a star-forming galaxy that happens to be the faintest and least luminous 
object of the twelve.  The measured slit-spectrum redshift confirms its location
within the void.

%\placefigure{fig:cone2}

The ELGs in the cone diagram appear to indicate the presence of a number of 
filaments and voids at these higher redshifts.  The ELGs map out large-scale structures, 
albeit only roughly, out to z = 0.09 without the need for follow-up spectra!  While the limited 
velocity resolution, together with the tendency for the ELGs to be less clustered,
prevents the use of these data for detailed definition of 
large-scale structures at these redshifts, they certainly provide a reasonable 
picture of the coarse outlines.  Furthermore, since the KISS galaxies are selected 
due to their strong line emission, follow-up spectroscopy of these faint galaxies 
is relatively easy.  Thus, using ELGs to trace out the main features in the spatial 
distribution of galaxies is an efficient use of telescope time.

\subsection{Comparison with Previous Surveys}

Table \ref{table:tab2} lists cross-references for KISS ELGs which are also cataloged 
in previous photographic surveys for active and star-forming galaxies.  
The first red KISS strip overlapped with four major surveys: Markarian (1967), Case 
(Pesch \& Sanduleak 1983), Wasilewski (1983), and UCM (Zamorano \etal 1994). 
As described in KR1, all ELGs cataloged by the emission line-selected surveys 
(Wasilewski and UCM) were recovered by KISS. The second red survey strip, however, 
overlaps with only two surveys, both of which select sources based on their UV-excess 
(Markarian) or on a combination of UV-excess and emission lines (Case). We expect KISS 
to recover only a subset of the objects cataloged by these surveys; in the first red 
KISS strip, 73\% of Markarian and Case objects were listed either in the main survey 
or among the 4$\sigma$ to 5$\sigma$ objects. A similar fraction of objects from
these catalogs was recovered in the second red survey strip, as detailed below.

There are 16 Markarian galaxies in the area of the second KISS survey strip; 
twelve of these (75\%) were recovered by KISS in the main survey (i.e., Table~\ref{table:tab2}).  
Of the four objects not found by KISS, one is beyond the redshift limit set by our 
survey filter and one has no redshift listed in the Markarian catalog. Inspection of 
the objective-prism spectra of the two remaining Markarian galaxies, Mrk 491 and 
Mrk 1475, revealed no evidence of line emission. 

There are 55 Case objects in the KISS area. For one of them (CG 646) it is noted in 
the NASA/IPAC Extragalactic Database\footnote{The NASA/IPAC Extragalactic Database 
(NED) is operated by the Jet Propulsion Laboratory, California Institute of Technology, 
under contract with the National Aeronautics and Space Administration.} 
that no object is present at the given position 
in the second Palomar Sky Survey images, and that this Case object may possibly have 
been a very faint dwarf nova.  We therefore do not consider CG 646 to be a {\it bona fide} 
galaxy. Of the remaining 54 galaxies, 39 were recovered by KISS in the main survey 
(72\%). We examined the 15 Case galaxies not recovered by KISS, and found that 
eight are listed in the Case Survey papers as being color-selected, and four are 
listed as having questionable (w? or w:) line detections. Only three of the 
unrecovered galaxies are listed as having strong or medium-strength emission lines. 
Careful examination of the KISS objective-prism spectra for these objects shows that
one galaxy (CG 501) appears to have emission lines, but they are too weak for it to 
be included in the KISS catalog.  The objective-prism spectra of the other two galaxies 
(CG 536 and CG 902) show no evidence of line emission.
Spectroscopy of 61 Case galaxies by Salzer \etal (1995) in the red portion of the 
spectrum found that 31\% had H$\alpha$ equivalent widths less than 30 \AA.  Hence, 
it would appear that the majority of non-detected Case galaxies belong to the 
weak-lined subsample of the Case survey, to which KISS is not sensitive.

%************************************************************************

\section{Summary}

We present the second list of H$\alpha$-selected emission-line galaxy candidates
(and third list overall) from the KPNO International Spectroscopic Survey (KISS).
All data presented here were obtained with the 0.61-m Burrell Schmidt telescope.
KISS is an objective-prism survey, but differs from older such surveys by virtue
of the fact that it utilizes a CCD as the detector.  While we sacrifice areal coverage
relative to classical photographic surveys, we benefit from the enormous gain in
sensitivity that CCDs provide over plates.  We readily detect strong-lined ELGs as 
faint as B = 21.  In addition, the pan-chromatic nature of CCDs allows us greater
wavelength agility compared to photographic surveys.  In combination with our
survey filter we are sensitive to a broader range of galaxian redshifts than the older
photographic objective-prism surveys (Paper I).  The combination of higher sensitivity, 
lower noise, and larger volumes surveyed yield huge improvements in the depth of the 
resulting survey.  KISS finds 169 times more AGN and starburst galaxy candidates 
per unit area than did the Markarian (1967) survey, and 30 times more than the UCM 
survey (Zamorano \etal 1994).

The current installment of KISS includes 1029 ELG candidates selected from 62 red 
survey fields covering a total of 65.8 deg$^2$.  This yields a surface density of 15.6 
galaxies per deg$^2$.   The two red survey lists combined (KR1 plus the current) cover an 
area of 128 square degrees with a surface density of nearly 17 galaxies per deg$^2$. 
We are sensitive to the H$\alpha$ emission line with redshifts up to $\sim$0.10.  Our
survey follows a narrow strip across the sky at a declination of $\delta$(1950) = 
43$\arcdeg$~30$\arcmin$ and spans the RA range 11$^h$~55$^m$ to 16$^h$~15$^m$. 
This region was chosen to pass through the center of the Bo\"otes void (Kirshner \etal 1981).  
For each object in the catalog we tabulate accurate equatorial coordinates, B \& V 
photometry, and estimates of the redshift and line strength measured from the 
objective-prism spectra. Also provided are finder charts and extracted spectral 
plots for all galaxies.  In addition to the main survey list, we include a 
supplementary list of 291 ELG candidates with weaker (lower significance) 
emission lines.  

One of the advantages of our survey method is the large amount of basic data
that we acquire for each object.  This in turn allows us to parameterize the 
constituents of the survey and to develop a fairly complete picture of the overall
sample without the need for extensive follow-up observations.  We present
an overview of the survey properties for the current list of ELG candidates.
The median apparent magnitude of the sample is B~=~18.13, which agrees
well with the value found for KR1 (B = 18.08), and which is substantially fainter 
than previous ELG surveys.  Objects fainter than B~=~20 are routinely cataloged.  
Line strengths measured from the objective-prism spectra show that KISS is 
sensitive to objects with H$\alpha$ + [\ion{N}{2}] equivalent widths of less than 
20 \AA, and that most objects with EW $>$ 30 \AA\ are detected.  The median 
emission-line flux of the KISS sample is more than three times lower than that of 
the UCM survey (Gallego \etal 1996).  The luminosity distribution of the KISS ELGs 
is heavily weighted toward intermediate- and low-luminosity galaxies, although we
are still sensitive to luminous AGN and starbursting galaxies.  The median absolute 
magnitude of M$_B$ = $-$18.67 underscores the fact that strong-lined galaxies of 
the type cataloged by KISS tend to be less luminous than the types of objects found 
in more traditional magnitude-limited samples.

While we stressed above that one can learn a great deal about each KISS ELG
from the survey data alone, it is still necessary to obtain higher dispersion
follow-up spectra in order to arrive at a more complete understanding of each object.
For example, the low-dispersion nature of the objective-prism spectra does not 
allow us to distinguish between AGN and star-formation activity in the KISS galaxies.
Further, the redshifts derived from the KISS spectral data are too coarse to be used 
in detailed spatial distribution studies (e.g., Lee, Salzer \& Gronwall 2004).  We are 
in the process of obtaining spectra for a large number of KISS ELGs from all three
survey lists (KR1, KB1, and the current list) in order to better assess the nature of the 
individual galaxies.  This in turn enables the sample to be used for a wide variety of
science applications, many of which are outlined in Paper I.  Examples include a
series of multi-wavelength studies of the properties of KISS ELGs in the radio
(Van Duyne \etal 2004) and X-rays (Stevenson \etal 2002), plus studies currently
underway in the far-IR (IRAS) and near-IR (2MASS).

\acknowledgments

We gratefully acknowledge financial support for the KISS project through
NSF Presidential Faculty Award to JJS (NSF-AST-9553020), which was instrumental
in allowing for the international collaboration, as well as continued support via
NSF-AST-0071114.  CG also acknowledges support from NSF-AST-0137927.  Several useful 
suggestions by the anonymous referee helped to improve the presentation of this paper.  
We thank the numerous colleagues with whom we have 
discussed the KISS project over the past several years, including Jes\'us Gallego, 
Rafael Guzm\'an, Rob Kennicutt, David Koo, Trinh Thuan, Alexei Kniazev, Yuri Izotov,
Janice Lee, Jason Melbourne and Jose Herrero.  Finally, we wish to thank Heather Morrison, 
Paul Harding, and the  Astronomy Department of Case Western Reserve University for 
maintaining the Burrell Schmidt during the period of time when the survey observations 
reported here were obtained.

%************************************************************************

\appendix
\section{Supplementary Table of 4$\sigma$ Objects}

As discussed in Section 3, our selection method has produced a list of objects
with apparent emission lines with strengths that are slightly weaker than the
5$\sigma$ limit imposed on the main survey objects.  Because one of the primary
goals of the KISS project is to construct a deep but statistically complete sample
of ELGs, we made the decision to exclude these objects from the main survey list.
However, these objects are nonetheless valid ELG candidates, and this list of
sources likely includes a number of interesting objects.  Therefore, rather than 
ignore these weaker-lined ELG candidates entirely, we are publishing them in
a supplementary table.

Listed in Table \ref{table:tab3} are 291 ELG candidates that have emission lines 
detected at between the 4$\sigma$ and 5$\sigma$ level.  The format of Table 
\ref{table:tab3} is the same as for Table \ref{table:tab2}, except that the objects 
are now labeled with KISSRx numbers (`x' for extra).  The KISSRx numbers start at 
190 since we presented 189 KISSRx objects in KR1. The full version of the table, 
as well as finder charts for all 291 KISSRx galaxies, are available in the electronic 
version of the paper. 

%\placetable{table:tab3}

The characteristics of the supplementary ELG sample are similar to those of the
main survey ELGs, although with some notable differences.  The median
H$\alpha$ equivalent width is 29.5 \AA, roughly a third lower than the value for 
the main sample.  The KISSRx galaxies are somewhat fainter (median B magnitude
of 18.77) and significantly redder (median B$-$V = 0.77).  Their median redshift
is slightly higher than that of the main sample (0.068), and their median
luminosity is slightly lower ($-$18.27).  Hence, the supplementary ELG list
appears to be dominated by intermediate luminosity galaxies with a significantly
lower rate of star-formation activity (lower equivalent widths, redder colors)
than the ELGs in the main sample.   The differences between the KISSR and KISSRx
objects in the current paper are similar to those seen between the two samples
in KR1.

%********************************* REFERENCES***************************

%\clearpage

% Now comes the reference list.  In this document, we used \cite to call
% out citations, so we must use \bibitem in the reference list, which
% means we use the LaTeX thebibliography environment.  Please note that
% \begin{thebibliography} is followed by a null argument.  If you forget
% this, mayhem ensues, and LaTeX will say "Perhaps a missing item?" when
% you run it.  Do not call us, do not send mail when this happens.  Put
% the silly {} after the \begin{thebibliography}.
%
% Each reference has a \bibitem command to define the citation format
% to be placed in the text (in []) and the symbolic tag used for 
% cross referencing (in {}).
%
% See sample1.tex, or the AASTeX guide, for an alternative to the \cite-
% \bibitem command.

%********************************* TABLE 1 ************************************

%\clearpage
%\renewcommand{\arraystretch}{.6}

\begin{deluxetable}{lccc}
%\scriptsize
%\rotate
\tablecolumns{4}
\tablewidth{0pt}
\tablecaption{KISS 43$^\circ$ Red Survey Observing Runs\label{table:tab1}}
\tablenum{1}
\tablehead{
\colhead{Dates of Run} & \colhead{Number of} & \colhead{Number of} & \colhead{Number of}\\
& \colhead{Nights\tablenotemark{a}} & \colhead{Fields -- Direct\tablenotemark{b}} & \colhead{Fields -- Spectral\tablenotemark{b}}\\
\colhead{(1)}&\colhead{(2)}&\colhead{(3)}&\colhead{(4)}
}
\startdata
April 19 -- May 4, 1998  & 15 & 43 & 40 \\
June 19 -- 21, 1998   & 3  & 7  & \nodata \\
May 7 -- 14, 1999    & 8  & 12 & 22 \\

\enddata
\tablenotetext{a}{Number of nights during run that data were obtained.}
\tablenotetext{b}{Number of survey fields observed.}
\end{deluxetable}

%****************************** TABLE 2 ***********************************

\clearpage

\tabletypesize{\scriptsize}
\renewcommand{\arraystretch}{.6}

\begin{deluxetable}{rrrllcccrrcl}
\scriptsize
%\rotate
\tablecolumns{12}
\tablewidth{0pt}
\tablecaption{List of Candidate ELGs\label{table:tab2}}
\tablenum{2}
\tablehead{
\colhead{KISSR}&\colhead{Field}&\colhead{ID}&\colhead{R.A.}&\colhead{Dec.}&\colhead{B}
&\colhead{B$-$V}&\colhead{z$_{KISS}$}&\colhead{Flux\tablenotemark{a}}
&\colhead{EW}&\colhead{Qual.}&\colhead{Comments}\\
\colhead{\#}&&&\colhead{(J2000)}&\colhead{(J2000)}&&&&&\colhead{[\AA]}&&\\
\colhead{(1)}&\colhead{(2)}&\colhead{(3)}&\colhead{(4)}
&\colhead{(5)}&\colhead{(6)}&\colhead{(7)}
&\colhead{(8)}&\colhead{(9)}&\colhead{(10)}&\colhead{(11)}&\colhead{(12)}
}
\startdata
 1129&F1155 & 5259&11 53 51.3&42 48 00.7&  20.84&   1.08&-0.0076&   66&   84& 2&\phm{imaveryveryverylongstring}\\
 1130&F1155 & 5256&11 54 08.8&43 23 22.5&  18.52&   0.50& 0.0440&   56&   71& 1&\phm{imaveryveryverylongstring}\\
 1131&F1155 & 5038&11 54 10.5&43 06 01.5&  18.13&   0.77& 0.0668&   91&   31& 1&\phm{imaveryveryverylongstring}\\
 1132&F1155 & 5324&11 54 10.5&43 32 54.2&  17.58&   0.56& 0.0662&  161&   67& 1&\phm{imaveryveryverylongstring}\\
 1133&F1155 & 5234&11 54 15.1&43 34 00.9&  16.77&   0.75& 0.0666&  485&   63& 1&\phm{imaveryveryverylongstring}\\
 1134&F1155 & 5003&11 54 15.7&43 13 36.1&  16.46&   0.57& 0.0157&  126&   12& 1& CG 1472                       \\
 1135&F1155 & 4840&11 54 18.7&43 06 57.4&  18.26&   0.46& 0.0656&   68&   64& 2&\phm{imaveryveryverylongstring}\\
 1136&F1155 & 4700&11 54 23.9&43 05 09.8&  15.94&   0.77& 0.0677&  530&   48& 1&\phm{imaveryveryverylongstring}\\
 1137&F1155 & 4797&11 54 25.1&43 16 59.2&  17.17&   0.69& 0.0543&  237&   47& 1&\phm{imaveryveryverylongstring}\\
 1138&F1155 & 4513&11 54 29.4&42 58 48.5&  16.30&   0.87& 0.0231&  217&   14& 1&\phm{imaveryveryverylongstring}\\
 \\
 1139&F1155 & 4526&11 54 33.7&43 08 12.8&  18.67&   0.50& 0.0867&  107&   97& 1&\phm{imaveryveryverylongstring}\\
 1140&F1155 & 4556&11 54 35.6&43 14 57.9&  16.34&   0.96& 0.0674&  252&   21& 1&\phm{imaveryveryverylongstring}\\
 1141&F1155 & 4280&11 54 45.0&43 09 54.0&  16.87&   0.92& 0.0678&  240&   31& 1&\phm{imaveryveryverylongstring}\\
 1142&F1155 & 3983&11 54 58.6&43 11 38.5&  19.39&   0.77& 0.0898&   41&   63& 2&\phm{imaveryveryverylongstring}\\
 1143&F1155 & 3513&11 55 32.0&43 31 29.1&  17.80&   0.47& 0.0349&  251&  118& 1&\phm{imaveryveryverylongstring}\\
 1144&F1155 & 2966&11 55 35.4&42 46 35.5&  19.56&   0.50& 0.0836&   66&  292& 2&\phm{imaveryveryverylongstring}\\
 1145&F1155 & 3534&11 55 35.8&43 40 34.7&  19.36&   0.54& 0.0499&   71&  253& 1&\phm{imaveryveryverylongstring}\\
 1146&F1155 & 3081&11 55 38.4&43 02 44.2&  14.59&   0.78& 0.0294&  788&   20& 1&UGC 6901, CG 1477              \\
 1147&F1155 & 3518&11 55 40.6&43 48 39.7&  17.08&   1.09& 0.0904&   55&    7& 1& CG 1478                       \\
 1148&F1155 & 3479&11 55 43.5&43 50 41.9&  17.91&   0.57& 0.0728&  255&  130& 1&\phm{imaveryveryverylongstring}\\
 \\
 1149&F1155 & 3531&11 55 50.4&44 08 06.6&  16.45&   0.94& 0.0734&  253&   31& 1&\phm{imaveryveryverylongstring}\\
 1150&F1155 & 3044&11 55 57.1&43 35 34.2&  15.30&   0.51& 0.0358&  333&   17& 1& Mk 1464, CG 1480              \\
 1151&F1155 & 3301&11 55 57.6&43 59 09.0&  19.22&   0.77& 0.0733&   33&   35& 3&\phm{imaveryveryverylongstring}\\
 1152&F1155 & 2836&11 56 12.0&43 45 10.1&  19.06&   0.44& 0.0486&   72&  122& 1&\phm{imaveryveryverylongstring}\\
 1153&F1155 & 2596&11 56 24.7&43 48 01.8&  18.27&   0.61& 0.0694&  129&  113& 1&\phm{imaveryveryverylongstring}\\
 1154&F1155 & 1892&11 56 32.9&42 59 38.8&  16.30&   0.76& 0.0747&  324&   46& 1&\phm{imaveryveryverylongstring}\\
 1155&F1155 & 1803&11 57 04.5&43 49 37.0&  16.34&   0.84& 0.0247&  344&   25& 1&\phm{imaveryveryverylongstring}\\
 1156&F1155 & 1159&11 57 15.1&43 15 58.4&  18.72&   0.69& 0.0901&   39&   34& 1&\phm{imaveryveryverylongstring}\\
 1157&F1155 & 1538&11 57 17.5&43 49 47.7&  17.50&   0.49& 0.0251&  189&   67& 1&\phm{imaveryveryverylongstring}\\
 1158&F1155 & 1282&11 57 32.1&43 57 16.1&  17.33&   0.73& 0.0704&  272&   59& 1&\phm{imaveryveryverylongstring}\\

\enddata
\tablenotetext{\,}{Note. -- The complete version of this table is presented in 
the electronic edition of the Journal.  A portion is shown here for guidance 
regarding its content and format.}
\tablenotetext{a}{Units of 10$^{-16}$ erg/s/cm$^2$}
\end{deluxetable}

%*************************** TABLE 3 ********************************

\clearpage

\begin{deluxetable}{ccccccccc}
\tablecaption{$V/V_{max}$ Test \label{table:comptab}}
\tablewidth{0pt }
\tabletypesize{\footnotesize}
\tablenum{3}
\tablehead{
\colhead{} &\colhead{Total}  &\colhead{Number} &\colhead{Number} &\colhead{}
&\colhead{Number} &\colhead{Cumulative} &\colhead{\%} &\colhead{\%}\\
\colhead{$m_L$} &\colhead{Number} &\colhead{Flux} &\colhead{Volume} 
&\colhead{$\langle V/V_{max}\rangle$} &\colhead {added}&\colhead{number}
&\colhead{Flux} &\colhead{Complete}\\
\colhead{} &\colhead{} &\colhead{Limited} &\colhead{Limited} &\colhead{}
&\colhead{} &\colhead{added} &\colhead{Limited}
&\colhead{}\\
\colhead{(1)}&\colhead{(2)}&\colhead{(3)}&\colhead{(4)}
&\colhead{(5)}&\colhead{(6)}&\colhead{(7)}
&\colhead{(8)}&\colhead{(9)}
}

\startdata
											
   13.0 & \phn\phn 17 & \phn 17 & \phn\phn 0 & 0.5485 & \phn 0 & \phn\phn 0 & 100.00 & 100.00\\
   13.1 & \phn\phn 21 & \phn 20 & \phn\phn 1 & 0.5794 & \phn 0 & \phn\phn 0 & \phn 95.24 & 100.00\\
   13.2 & \phn\phn 26 & \phn 25 & \phn\phn 1 & 0.5906 & \phn 0 & \phn\phn 0 & \phn 96.15 & 100.00\\
   13.3 & \phn\phn 30 & \phn 28 & \phn\phn 2 & 0.5687 & \phn 0 & \phn\phn 0 & \phn 93.33 & 100.00\\
   13.4 & \phn\phn 34 & \phn 32 & \phn\phn 2 & 0.5520 & \phn 0 & \phn\phn 0 & \phn 94.12 & 100.00\\
   13.5 & \phn\phn 42 & \phn 39 & \phn\phn 3 & 0.5656 & \phn 0 & \phn\phn 0 & \phn 92.86 & 100.00\\
   13.6 & \phn\phn 47 & \phn 42 & \phn\phn 5 & 0.5499 & \phn 0 & \phn\phn 0 & \phn 89.36 & 100.00\\
   13.7 & \phn\phn 55 & \phn 46 & \phn\phn 9 & 0.5382 & \phn 0 & \phn\phn 0 & \phn 83.64 & 100.00\\
   13.8 & \phn\phn 71 & \phn 58 & \phn 13 & 0.5843 & \phn 0 & \phn\phn 0 & \phn 81.69 & 100.00\\
   13.9 & \phn\phn 75 & \phn 60 & \phn 15 & 0.5258 & \phn 0 & \phn\phn 0 & \phn 80.00 & 100.00\\
   14.0 & \phn\phn 87 & \phn 65 & \phn 22 & 0.5582 & \phn 0 & \phn\phn 0 & \phn 74.71 & 100.00\\
   14.1 & \phn 105 & \phn 79 & \phn 26 & 0.5777 & \phn 0 & \phn\phn 0 & \phn 75.24 & 100.00\\
   14.2 & \phn 128 &  102 & \phn 26 & 0.5993 & \phn 0 & \phn\phn 0 & \phn 79.69 & 100.00\\
   14.3 & \phn 150 &  116 & \phn 34 & 0.5932 & \phn 0 & \phn\phn 0 & \phn 77.33 & 100.00\\
   14.4 & \phn 177 &  138 & \phn 39 & 0.5889 & \phn 0 & \phn\phn 0 & \phn 77.97 & 100.00\\
   14.5 & \phn 209 &  157 & \phn 52 & 0.5903 & \phn 0 & \phn\phn 0 & \phn 75.12 & 100.00\\
   14.6 & \phn 238 &  175 & \phn 63 & 0.5768 & \phn 0 & \phn\phn 0 & \phn 73.53 & 100.00\\
   14.7 & \phn 277 &  193 & \phn 84 & 0.5757 & \phn 0 & \phn\phn 0 & \phn 69.68 & 100.00\\
   14.8 & \phn 308 &  213 & \phn 95 & 0.5528 & \phn 0 & \phn\phn 0 & \phn 69.16 & 100.00\\
   14.9 & \phn 346 &  236 &  110 & 0.5444 & \phn 0 & \phn\phn 0 & \phn 68.21 & 100.00\\
   15.0 & \phn 393 &  256 &  137 & 0.5480 & \phn 0 & \phn\phn 0 & \phn 65.14 & 100.00\\
   15.1 & \phn 431 &  262 &  169 & 0.5267 & \phn 0 & \phn\phn 0 & \phn 60.79 & 100.00\\
   15.2 & \phn 476 &  276 &  200 & 0.5192 & \phn 0 & \phn\phn 0 & \phn 57.98 & 100.00\\
   15.3 & \phn 520 &  302 &  218 & 0.5108 & \phn 0 & \phn\phn 0 & \phn 58.08 & 100.00\\
   15.4 & \phn 579 &  323 &  256 & 0.5122 & \phn 0 & \phn\phn 0 & \phn 55.79 & 100.00\\
   15.5 & \phn 630 &  340 &  290 & 0.5039 & \phn 0 & \phn\phn 0 & \phn 53.97 & 100.00\\
   15.6 & \phn 683 &  364 &  319 & 0.4983 & \phn 1 & \phn\phn 1 & \phn 53.29 & \phn 99.73\\
   15.7 & \phn 728 &  371 &  357 & 0.4824 &   12 & \phn 13 & \phn 50.96 & \phn 96.61\\
   15.8 & \phn 773 &  388 &  385 & 0.4745 &   10 & \phn 23 & \phn 50.19 & \phn 94.40\\
   15.9 & \phn 815 &  403 &  412 & 0.4577 &   20 & \phn 43 & \phn 49.45 & \phn 90.36\\
   16.0 & \phn 859 &  405 &  454 & 0.4423 &   23 & \phn 66 & \phn 47.15 & \phn 85.99\\
   16.1 & \phn 901 &  416 &  485 & 0.4264 &   29 & \phn 95 & \phn 46.17 & \phn 81.41\\
   16.2 & \phn 923 &  388 &  535 & 0.3900 &   44 &  139 & \phn 42.04 & \phn 73.62\\
   16.3 & \phn 955 &  380 &  575 & 0.3719 &   41 &  180 & \phn 39.79 & \phn 67.86\\
   16.4 & \phn 977 &  359 &  618 & 0.3571 &   41 &  221 & \phn 36.75 & \phn 61.90\\
   16.5 & \phn 998 &  345 &  653 & 0.3373 &   51 &  272 & \phn 34.57 & \phn 55.92\\
   16.6 & 1008 &  332 &  676 & 0.3106 &   64 &  336 & \phn 32.94 & \phn 49.70\\
   16.7 & 1014 &  317 &  697 & 0.2769 &   75 &  411 & \phn 31.26 & \phn 43.54\\
   16.8 & 1016 &  296 &  720 & 0.2416 &   83 &  494 & \phn 29.13 & \phn 37.47\\
   16.9 & 1018 &  268 &  750 & 0.2164 &   82 &  576 & \phn 26.33 & \phn 31.75\\
   17.0 & 1019 &  244 &  775 & 0.1941 &   91 &  667 & \phn 23.95 & \phn 26.78\\

\enddata
\end{deluxetable}

%*************************** TABLE 4 ********************************

\clearpage

\tabletypesize{\scriptsize}
\renewcommand{\arraystretch}{.6}

\begin{deluxetable}{rrrllcccrrcl}
\scriptsize
\tablecolumns{12}
\tablewidth{0pt}
\tablecaption{List of 4$\sigma$ Candidate ELGs\label{table:tab3}}
\tablenum{4}
\tablehead{
\colhead{KISSRx}&\colhead{Field}&\colhead{ID}&\colhead{R.A.}&\colhead{Dec.}&\colhead{B}
&\colhead{B$-$V}&\colhead{z$_{KISS}$}&\colhead{Flux\tablenotemark{a}}
&\colhead{EW}&\colhead{Qual.}&\colhead{Comments}\\
\colhead{\#}&&&\colhead{(J2000)}&\colhead{(J2000)}&&&&&\colhead{[\AA]}&&\\
\colhead{(1)}&\colhead{(2)}&\colhead{(3)}&\colhead{(4)}
&\colhead{(5)}&\colhead{(6)}&\colhead{(7)}
&\colhead{(8)}&\colhead{(9)}&\colhead{(10)}&\colhead{(11)}&\colhead{(12)}
}
\startdata
  190&F1155 & 5353&11 54 10.0&43 33 31.1&  18.45&   0.46& 0.0671&   27&   28& 3&\phm{imaveryveryverylongstring}\\
  191&F1155 & 4820&11 54 23.2&43 14 35.6&  18.33&   0.81& 0.0879&   39&   17& 3&\phm{imaveryveryverylongstring}\\
  192&F1155 & 4935&11 54 30.9&43 38 55.9&  17.78&   0.98& 0.0890&   21&    5& 3&\phm{imaveryveryverylongstring}\\
  193&F1155 & 3506&11 55 35.7&43 38 07.4&  18.67&   0.74& 0.0746&  106&  115& 2&\phm{imaveryveryverylongstring}\\
  194&F1155 & 3327&11 55 54.9&43 56 32.8&  16.08&   0.54& 0.0246&  165&   21& 2&\phm{imaveryveryverylongstring}\\
  195&F1155 & 2903&11 56 14.0&43 55 18.8&  18.64&   0.53& 0.0774&   39&   48& 3&\phm{imaveryveryverylongstring}\\
  196&F1155 & 1846&11 57 00.9&43 47 43.7&  19.13&   0.63& 0.0480&   52&   67& 3&\phm{imaveryveryverylongstring}\\
  197&F1155 &  256&11 57 37.7&42 43 19.9&  18.00&   0.84& 0.0738&   59&   24& 2&\phm{imaveryveryverylongstring}\\
  198&F1200 & 5996&11 58 28.5&43 04 31.7&  16.68&   0.62& 0.0173&   36&    5& 3&\phm{imaveryveryverylongstring}\\
  199&F1200 & 2782&12 00 09.8&44 08 40.0&  18.86&   0.63& 0.0836&   56&   89& 3&\phm{imaveryveryverylongstring}\\
 \\
  200&F1200 & 1587&12 00 58.5&44 09 05.4&  18.04&   0.69& 0.0035&   52&   20& 2&\phm{imaveryveryverylongstring}\\
  201&F1200 & 1639&12 01 15.4&43 19 09.3&  19.62&   0.85& 0.0809&   23&   73& 3&\phm{imaveryveryverylongstring}\\
  202&F1200 & 1157&12 01 36.1&43 17 23.6&  18.67&   1.05& 0.0692&   45&   19& 2&\phm{imaveryveryverylongstring}\\
  203&F1204 & 6213&12 02 53.5&42 52 35.4&  20.06&   1.03& 0.0927&   56&   81& 2&\phm{imaveryveryverylongstring}\\
  204&F1204 & 5164&12 02 59.6&44 04 07.1&  21.30&   2.01& 0.0726&   31&   41& 2&\phm{imaveryveryverylongstring}\\
  205&F1204 & 4187&12 03 48.8&43 28 28.6&  17.94&   0.81& 0.0761&   80&   42& 2&\phm{imaveryveryverylongstring}\\
  206&F1204 & 2054&12 05 31.4&42 39 01.2&  19.34&   0.69& 0.0544&   41&   66& 2&\phm{imaveryveryverylongstring}\\
  207&F1208 & 4425&12 07 53.7&43 20 50.0&  18.62&   0.90& 0.0667&   56&   35& 2&\phm{imaveryveryverylongstring}\\
  208&F1208 & 3147&12 08 52.8&42 51 34.7&  20.44&   1.65& 0.0728&   55&   34& 2&\phm{imaveryveryverylongstring}\\
  209&F1212 & 2565&12 13 16.8&44 09 25.4&  20.39&   1.86& 0.0831&   21&   24& 3&\phm{imaveryveryverylongstring}\\
 \\
  210&F1212 & 2767&12 13 27.7&43 22 50.7&  20.77&   0.71& 0.0682&   46&  430& 3&\phm{imaveryveryverylongstring}\\
  211&F1212 & 1358&12 14 02.2&43 52 17.1&  19.30&   1.10& 0.0681&   35&   33& 3&\phm{imaveryveryverylongstring}\\
  212&F1212 &  794&12 14 22.3&43 53 45.5&  19.15&   0.51& 0.0922&   47&   76& 3&\phm{imaveryveryverylongstring}\\
  213&F1212 &  146&12 15 07.0&42 54 47.7&  18.46&   0.94& 0.0600&   42&   23& 3&\phm{imaveryveryverylongstring}\\
  214&F1217 & 4382&12 16 42.9&43 40 41.8&  17.76&   0.59& 0.0412&   38&   13& 2&\phm{imaveryveryverylongstring}\\
  215&F1217 & 3456&12 17 36.7&42 48 30.6&  18.43&   0.62& 0.0868&   48&   48& 2&\phm{imaveryveryverylongstring}\\
  216&F1217 & 1873&12 18 08.2&43 53 24.6&  18.97&   0.83& 0.0659&   19&   19& 3&\phm{imaveryveryverylongstring}\\
  217&F1217 & 1750&12 18 23.5&43 24 55.9&  18.16&   0.48& 0.0449&   26&   18& 2&\phm{imaveryveryverylongstring}\\
  218&F1221 & 6045&12 19 42.2&43 52 10.5&  16.92&   0.53& 0.0499&   46&   14& 2&\phm{imaveryveryverylongstring}\\
  219&F1221 & 3278&12 21 26.7&43 52 56.0&  17.62&   0.75& 0.0454&   82&   25& 1&\phm{imaveryveryverylongstring}\\
\enddata
\tablenotetext{\,}{Note.--- The complete version of this table is presented in the
electronic edition of the Journal.  A portion is shown here for guidance regarding
its content and format.}
\tablenotetext{a}{Units of 10$^{-16}$ erg/s/cm$^2$}
\end{deluxetable}


\begin{thebibliography}{}

%\bibitem[Burstein \& Heiles 1984]{bh84} Burstein, D., \& Heiles, C.  1984, \apjs, 54, 33

\bibitem[Aldering 1990]{ga90} Aldering, G. S.  1990, PhD Thesis, University of Michigan

\bibitem[Falco \etal 1999]{falco99} Falco, E. E., Kurtz, M. J., Geller, M. J., 
Huchra, J. P., Peters, J., Berlind, P., Mink, D. J., Tokarz, S. P., \& Elwell, B. 
1999, \pasp, 111, 438

\bibitem[Fukugita \etal 1995]{FSI95} Fukugita, M., Shimasaku, K., \& Ichikawa, T.  
1995, \pasp, 107, 945

\bibitem[Gallego \etal 1996]{JG96} Gallego, J., Zamorano, J., Rego, M., Alonso, O.,
\& Vitores, A. G. 1996, \aaps, 120, 323

%\bibitem[Geller \etal 1997]{geller97} Geller, M. J., Kurtz, M. J., Wegner, G.,
%Thorstensen, J. R., Fabricant, D. G., Marzke, R. O., Huchra, J. P., Schild, R. E.,
%\& Falco, E. E.  1997, \aj, 114, 2205

\bibitem[Gronwall \etal 2004]{CG2} Gronwall, C.,  Salzer, J. J., Brenneman, L., Condy, E., \& Santos, M.
2004, in preparation

%\bibitem[Gronwall \etal 2004]{CGspec} Gronwall, C., Salzer, J. J., McKinstry, K., \& Wegner, G.
%2004, in preparation

%\bibitem[Herrero \etal 2000]{herrero00} Herrero, J. L., Frattare, L. M., Salzer, J. J., Gronwall, C., 
%\& Kearns, K.  2000, \pasp, submitted

\bibitem[Huchra \& Sargent(1973)]{huchra} Huchra, J. P., \& Sargent, W. L. W. 
1973, \apj, 186, 433

\bibitem[Kirshner \etal 1981]{kirsh81} Kirshner, R. P., Oemler, A., Jr., Schechter, P. L., 
\& Shectman, S. A.  1981, \apj, 248, L57

\bibitem[Kirshner \etal 1983]{kirsh83} Kirshner, R. P., Oemler, A., Jr., Schechter, P. L., 
\& Shectman, S. A.
1983, in {\it IAU Symposium 104, Early Evolution of the Universe and Its Present Structure}, 
eds. G. O. Abell and G. Chincarini (Dordrecht: Reidel), p. 97

\bibitem[Kirshner \etal 1987]{kirsh87} Kirshner, R. P., Oemler, A., Jr., Schechter, P. L., 
\& Shectman, S. A.  1987, \apj, 314, 493

\bibitem[Lee, Salzer \& Gronwall 2004]{LSG} Lee, J. C., Salzer, J. J., \& Gronwall, C. 
2004, in preparation

\bibitem[Lee \etal 2000]{LSLR} Lee, J. C., Salzer, J. J., Law, D. A., \& Rosenberg, J. L.
2000, \apj, 536, 606

\bibitem[MacAlpine \etal 1977]{Mac77} MacAlpine, G. M., Smith, S. B., 
\& Lewis, D. W. 1977, \apjs, 34, 95

\bibitem[Markarian \etal 1967]{Mrk67} Markarian, B. E. 1967, Astrofizika, 3, 55

\bibitem[Markarian \etal 1983]{Mrk83} Markarian, B. E., Lipovetskii, V. A., 
\& Stepanian, D. A. 1983, Astrofizika, 19, 29

\bibitem[Mazzarella \& Balzano 1986]{MB86} Mazzarella, J. M., \& Balzano, V. A. 
1986, \apjs, 62, 751

\bibitem[Moody \etal 1987]{Moody87} Moody, J. W., Kirshner, R. P., MacAlpine, G. M., \& 
Gregory, S. A.  1987, \apjl, 314, 33

\bibitem[Nilson 1973]{UGC} Nilson, P.  1973, {\it Uppsala General Catalogue of Galaxies}, 
(Uppsala: Roy. Soc. Sci. Uppsala)

\bibitem[P\'erez-Gonz\'alez \etal 2000]{UCMphot} P\'erez-Gonz\'alez, P. G., Zamorano, J., 
Gallego, J., \& Gil de Pez, A.  2000, \aaps, 141, 409

\bibitem[Pesch \& Sanduleak 1983]{Case83} Pesch, P., \& Sanduleak, N. 1983, 
\apjs, 51, 171

\bibitem[Popescu \etal 1997]{Pop97} Popescu, C. C., Hopp, U., \&  Els\"asser, H.  
1997, \aap, 328, 756

\bibitem[Popescu \etal 1996]{Ham96} Popescu, C. C., Hopp, U., Hagen, H. J., \&
Els\"asser, H.  1996, \aaps, 116, 43

\bibitem[Pustil'nik \etal 1995]{P95} Pustil'nik, S. Ugryumov, A. V., Lipovetsky, V. A.,
Thuan, T. X., \& Guseva, N. 1995, \apj, 443, 499

\bibitem[Roberts \& Haynes 1994]{RH94} Roberts, M. S., \& Haynes, M. P. 1994, \araa, 32, 115

\bibitem[Rosenberg \etal 1994]{R94} Rosenberg, J. L., Salzer, J. J., \& Moody, J. W.  
1994, \aj, 108, 1557

\bibitem[Salzer 1989]{S89} Salzer, J. J. 1989, \apj, 347, 152

\bibitem[Salzer \etal 2000]{kiss1} Salzer, J. J., Gronwall, C., 
Lipovetsky, V. A., Kniazev, A., Moody, J. W., Boroson, T. A., Thuan, T. X.,
Izotov, Y. I., Herrero, J. L., \& Frattare, L. M.  2000, \aj, 120, 80 (Paper I)

\bibitem[Salzer \etal 2001]{kissred1} Salzer, J. J., Gronwall, C., 
Lipovetsky,  V. A., Kniazev, A., Moody, J. W., Boroson, T. A., Thuan, T. X.,
Izotov, Y. I., Herrero, J. L., \& Frattare, L. M.  2001, \aj, 121, 66 (KR1)

\bibitem[Salzer \etal 2002]{kissblue1} Salzer, J. J., Gronwall, C., Sarajedini, V. L., 
Lipovetsky,  V. A., Kniazev, A., Moody, J. W., Boroson, T. A., Thuan, T. X.,
Izotov, Y. I., Herrero, J. L., \& Frattare, L. M.  2002, \aj, 123, 1292 (KB1)

\bibitem[Salzer \etal 1989]{SMB89} Salzer, J. J., MacAlpine, G. M., \& Boroson, 
T. A. 1989, \apjs, 70, 479

\bibitem[Salzer \etal 1995]{S95} Salzer, J. J., Moody, J. W., Rosenberg, J. L.,
Gregory, S. A., \& Newberry, M. V.  1995, \aj, 109, 2376

\bibitem[Sanduleak \& Pesch 1982]{Case82} Sanduleak, N. \& Pesch, P. 1982, \apjl, 258, 11

\bibitem[Sanduleak \& Pesch 1982]{Case87} Sanduleak, N. \& Pesch, P. 1987, \apjs, 63, 809

\bibitem[Schechter 1976]{Sch76} Schechter, P. L. 1976, \apj, 203, 297

\bibitem[Schlegel \etal 1998]{Sch98} Schlegel, D. J., Finkbeiner, D. P., \& Davis, M. 
1998, \apj, 500, 525

\bibitem[Schmidt(1968)]{schmidt} Schmidt, M. 1968, \apj, 151, 393

\bibitem[Smith \etal 1976]{Tol76} Smith, M. G., Aguirre, C., \& Zemelman, M. 
1976, \apjs, 32, 217

\bibitem[Stevenson \etal 2002]{Sam02}Stevenson, S. L., Salzer, J. J., Sarajedini, V. L., \&
Moran, E. C.  2002, \aj, 124, 3465

\bibitem[Surace \& Comte 1998]{Mar98} Surace, C., \& Comte, G.  1998, \aaps, 133, 171

\bibitem[Ugryumov \etal 1999]{Ugry99} Ugryumov, A. V., \etal 1999, \aaps, 135, 511

\bibitem[Van Duyne \etal 2004]{JVD04} Van Duyne, J., Beckerman, E., Salzer, J. J., Gronwall, C.,
Thuan, T. X., Condon, J. J., \& Frattare, L. M.  2004, \aj, submitted

\bibitem[Wasilewski 1983]{Was83} Wasilewski, A. J. 1983, \apj, 272, 68

\bibitem[Wegner \etal 2003]{Weg03} Wegner, G., Salzer, J. J., Jangren, A., Gronwall, C., \&
Melbourne, J.  2003, \aj, 125, 2373

\bibitem[Zamorano \etal 1994]{UCM92} Zamorano, J., Rego, M., Gallego, J., 
Vitores, A.G., Gonz\'alez-Riestra, R., \& Rodr\'iguez-Caderot, G.  
1994, \apjs, 95, 387

\bibitem[Zwicky \etal. 1961]{CGCG} Zwicky, F., Herzog, E., Kowal, C. T., Wild, P.,
\& Karpowicz, M.  1961--1968, {\it Catalogue of Galaxies and Clusters of Galaxies}, 
(Pasadena: CIT) (CGCG)

\end{thebibliography}
\end{document}